\documentclass[reprint,unsortedaddress,nofootinbib, amsmath,amssymb,aps,prd,showkeys]{revtex4-1}
\usepackage{subfig}
\usepackage{graphicx}
\usepackage{dcolumn}
\usepackage{bm}
\usepackage{hyperref}
\usepackage{physics}
\usepackage{xcolor}
\usepackage[maxfloats=256]{morefloats}
\usepackage{ragged2e}
\usepackage{comment} 
\usepackage{dcolumn}
\usepackage{bm}%
\usepackage{float}	
\usepackage[T1]{fontenc} 
\usepackage{color}
\usepackage{mdframed}

\usepackage{textcomp}

\usepackage{enumerate}
\usepackage{mathrsfs}
\usepackage{verbatim}
\usepackage{slashed}
\usepackage{epsfig}
\usepackage{enumerate}
\usepackage{wrapfig}
\usepackage{amsfonts}
\usepackage{multirow}
\usepackage{tikz}
\usepackage{amsmath}
\usepackage[normalem]{ulem}
\usepackage[normalem]{ulem}

\newcommand{\bea}{\begin{eqnarray}}
\newcommand{\eea}{\end{eqnarray}}
\newcommand{\be}{\begin{equation}}
	\newcommand{\ee}{\end{equation}}
\newcommand{\bal}{\begin{align}}
\newcommand{\ealign}{\end{align}}
\newcommand{\eal}{\end{align}}
\newcommand{\ben}{\begin{enumerate}}
\newcommand{\een}{\end{enumerate}}
\newcommand{\bit}{\begin{itemize}}
	
\newcommand{\nn}{\nonumber}

\newcommand{\bd}{\begin{displaymath}}
\newcommand{\ed}{\end{displaymath}}

\def\l{\lambda}

\newcommand{\mhpm}{\ensuremath{m_{H^{\pm}} \,\, }}

\newcommand{\vsm}{\ensuremath{v_{_{\! \rm SM}}} }
\newcommand{\mw}{\ensuremath{m_{_W} }}
\newcommand{\mwsm}{\ensuremath{m_{_W}^{\rm SM} }}
\newcommand{\mwcdf}{\ensuremath{m_{_W}^{\rm CDF} }}
\newcommand{\mwcms}{\ensuremath{m_{_W}^{\rm CMS} }}
\newcommand{\mz}{\ensuremath{m_{_Z} }}

\begin{document}


\title{$W-$mass and Muon $g-2$ in   Inert 2HDM  Extended by Singlet Complex Scalar}
\author{Hrishabh Bharadwaj}
\email{hrishabhphysics@gmail.com}
\affiliation{Rajkiya Mahila Mahavidyalaya, Budaun, Mahatmaa Jyotiba Phule Rohilkhand University, Uttar Pradesh, India}
\author{Mamta Dahiya}
\email{mamta.phy26@gmail.com}
\email{mamta.dahiya@sgtbkhalsa.du.ac.in}
\affiliation{SGTB Khalsa College, University of Delhi, Delhi, India.}
\author{Sukanta Dutta}
\email{Sukanta.Dutta@gmail.com}
\email{sukanta.dutta@sgtbkhalsa.du.ac.in}
\affiliation{SGTB Khalsa College, University of Delhi, Delhi, India.}
\affiliation{Delhi School of Analytics, Institution of Eminence, University of Delhi, Delhi.}
\author{Ashok Goyal}
\email{agoyal45@yahoo.com}
\affiliation{Department of Physics $\&$ Astrophysics, University of Delhi, Delhi, India.}
\begin{abstract}
The deviations of the recent measurements of the muon magnetic moment and the $W-$boson mass from their SM predictions  hint to new physics beyond the SM.  In this article, we address the observed discrepancies in the $W$-boson mass  and muon anomalous magnetic moment in  the Inert Two Higgs Doublet Model (I2HDM) extended by a  complex scalar field singlet under the SM gauge group. The model is constrained from the existing LEP data and the measurements of partial decay widths to gauge bosons at LHC.  It is shown that a large subset of the constrained  parameter space of the model  can  accommodate both, the experimentally measured  as well as SM global fit value of $W$-boson mass  while simultaneously explaining the observed muon  $g-2$ anomaly.
\end{abstract}
\maketitle

\section{Introduction}
\label{sec:intro}
The departures of  low-energy observables from their Standard Model (SM) predictions can provide indirect clues for physics beyond the SM.  The disappearing observed $W$ boson mass anomaly and the prevailing discrepancy in anomalous magnetic moment of muon  provide a stringent test
of the SM~\cite{Crivellin:2023zui} and should be explained by any proposed model  beyond SM. 

Until its recent measurement by CMS collaboration~\cite{CMS:2024nau}, the most precise known value  of the mass of  $W$  boson \mw was 
\bea
\mwcdf =  (80.4335 \pm 0.0094)\, {\text{GeV}},
\label{eq:wmass-cdf}
\eea
a measurement done by the CDF Collaboration~\cite{CDF:2022hxs}   on their full Run-2 dataset of  8.8 fb$^{-1}$.
This value deviates from the global average of the other experiments\cite{LHC-TeVMWWorkingGroup:2023zkn,Workman:2022ynf}
\bea
\mw^{\rm PDG} = (80.377 \pm 0.012)\,  (80.3692 \pm 0.0133)\,  {\text{GeV}}. 
\label{eq:wmass-av}
\eea
\noindent 
A global fit to electroweak data, used to predict \mw in the standard model, yields the 
value~\cite{deBlas:2022hdk} 
\bea
\mwsm =  (80.3499 \pm 0.0094 )\, {\text{GeV}}  
\label{eq:wmass-SM}
\eea
which is about $7\sigma$  below the value reported by CDF. 
Such a significant discrepancy, calls for a thorough investigation of physics beyond the Standard Model (BSM)~\cite{Kotwal:2024ume}.

However, recently, the CMS Collaboration has  reported their  first  measurement  of $W$ mass~\cite{CMS:2024nau} 
\bea
\mwcms =  (80.3602 \pm 0.0099)\, {\text{GeV}},
\label{eq:wmass-cms}
\eea
with a precision very similar to that of the recent CDF measurement~\cite{CDF:2022hxs} and better than that of all previous results. 
This value of  $W$ mass not is consistent with the expectation from the SM electroweak fit within experimental uncertainties as well as the present world average (excluding CDF). However, the the CDF measurement is way above this value.

Another  long standing  discrepancy is in the muon anomalous magnetic moment   where the direct measurements of muon ($g-2$) are precisely made and have been confirmed in several experiments \cite{Keshavarzi:2021eqa} .  The most recent measurement of the anomalous  muon magnetic moment by 
the Fermilab Muon $g -2$ Experiment~\cite{Muong-2:2023cdq} using data collected in 2019 and 2020 gives 
\bea
a_{\mu} &=&  \frac{(g-2)_\mu}{2}  = 116592057(25) \times 10^{-11} (0.21\, \text{ppm} )\nn\\
\label{amu-Muong}
\eea
resulting in the new world average 
\bea
a_{\mu}^{\rm exp}  = 116592059(22) \times 10^{-11} (0.19\, \text{ppm} )\label{amu-av}
\eea
The SM prediction is given by~\cite{AOYAMA20201}
\bea
a_{\mu}^{\rm SM} &=&   116591810(43)  \times 10^{-11}\label{amu-sm}
\eea
amounting to about $5\sigma$ discrepancy
\bea
\Delta a_{\mu} &=& a_{\mu}^{\rm exp}  -   a_{\mu}^{\rm SM} 
=\left(2.49 \pm 0.48  \right) \times 10^{-9}. \label{eq:delamu}
\eea
This SM prediction uses the conservative  leading order data-driven computation of Hadronic Vacuum  Polarisation (HVP)~\cite{Keshavarzi:2019abf} based on  the available data sets for the $ e^+ e^- \to  $hadrons cross section and the techniques applied for the evaluation of the HVP dispersive integral.  There is however, tension between the results of hadron vacuum polarisation from Lattice simulations of QCD~\cite{Boccaletti:2024guq, Kuberski:2023qgx}. This and the recent measurements of $e^+e^-$ going to hadrons by CMD3 Collaboration will make SM predictions closer to the experimental measurements.
The  prospects for improvements of the uncertainties in the SM prediction~\cite{AOYAMA20201} 
may make it closer to the experimental measurements\cite{CMD-3:2023alj, PhysRevD.108.054507}. However, how would this discrepancy play out by future analysis is not yet settled. 

The additional quantum corrections induced by new particles in a  model beyond SM might  account for the  observed anomaly in the $W-$boson mass as well as the muon magnetic moment.  These twin problems
have been addressed recently (either individually or simultaneously) in many models beyond the SM~\cite{Cici:2024fsl, *Borah:2023hqw,*Zhu:2023qyt, *Davoudiasl:2023cnc, *Abdallah:2023pbl, *Ahmadvand:2023gse, *Chakrabarty:2022gqi, *Bandyopadhyay:2022bgx,*Abdallah:2022shy, *Kim:2022axk,  *Zheng:2022ssr, *Chakrabarty:2022voz, *Kawamura:2022fhm, *Chowdhury:2022dps, *He:2022zjz, *Kim:2022xuo,  *Botella:2022rte, *Bhaskar:2022vgk,*Arcadi:2022dmt, *Kawamura:2022uft, *Zhou:2022cql, *Athron:2022qpo, *Babu:2022pdn, *Bagnaschi:2022qhb} with varying degrees of success. 

In an earlier work  \cite{Bharadwaj:2021tgp}, the authors have  addressed the observed discrepancies in the anomalous magnetic moment of muons and electrons by I2HDM to include a complex scalar field and a charged singlet vector-like lepton. In this spirit we revisit our earlier model albeit without the introduction of a charged vector-like lepton and discuss the constraints on the model parameters from the LEP data and recent Higgs decay data. Using this constrained model, we attempt to  address the possibility of explaining the observed  upward pull for $m_W$ and muon $g-2$. 

The rest of this  article is organised as follows:  The section~\ref{sec:model} briefly reviews our model. In section \ref{sec:constraints}, we discuss the constraints on model parameters coming from the Higgs decay and the LEP data.  The additional contributions to muon anomalous magnetic moment and  $W-$mass in our model are discussed in section~\ref{sec:mdm-mw}. The corresponding numerical results of the regions in parameter space that simultaneously satisfy the experimental results of both observables, namely the $W$ mass and the muon $g-2$ are given in section~\ref{sec:results}. In the end, we summarise our results in the section~\ref{sec:summary}.
\section{The model}
\label{sec:model}
The I2HDM consiss of two $SU(2)_L$ doublets of complex scalar fields: SM-like doublet 
$\Phi_1$ and another doublet $\Phi_2$(the inert doublet)  possessing the same
quantum numbers as  $\Phi_1$  but with no direct coupling to fermions.
We consider  a model with the scalar sector of 
I2HDM extended by a neutral complex gauge singlet scalar field 
$\Phi_3$. After electroweak symmetry breaking (EWSB),  $\Phi_1$  as well as $\Phi_3$ acquire   nonzero  real  vacuum expectation values, $\vsm$ and  $v_s$ respectively.   We invoke a $Z_2$ symmetry under which all SM fields  and 
$\Phi_1$ are even. The inert doublet fields $\Phi_2$  and the singlet scalar $\Phi_3$ are odd under this $Z_2$ symmetry.
Due to this symmetry the scalar fields in $\Phi_2$ do not mix with the SM-like field from $\Phi_1$.
The $Z_2$ symmetry  also ensures that the SM gauge bosons and fermions are forbidden to have direct interaction with the inert doublet and additional complex scalar singlet.  We however, allow an explicit breaking of $Z_2$ symmetry in the Yukawa Lagrangian ${\cal L}_Y$ in order to facilitate coupling of SM leptons with $CP$ odd pseudoscalars. 
\par   The part of the Lagrangian different from SM Lagrangian is written as
\begin{subequations}
	\begin{widetext}
		\bea
		{\cal L}&\supset& {\cal L}_{\rm scalar}\ +\ {\cal L}_{\rm Yukawa}\ 
		\eea
		\bea
		{\cal L}_{\rm scalar}&=& \left(D_\mu\Phi_1\right)^\dagger\ \left(D^\mu\Phi_1\right)\ +\ \left(D_\mu\Phi_2\right)^\dagger\ \left(D_\mu\Phi_2\right)\ +\ \left(D_\mu\Phi_3\right)^*\ \left(D_\mu\Phi_3\right)\ -\ V_{\rm scalar} \label{eq:lag-scalar}\\
		V_{\rm scalar}&=& V_{\rm 2 HDM}\left(\Phi_1, \Phi_2\right)+ V_{\rm Singlet} \left(\Phi_3\right) +  V_{\rm Mix}\left(\Phi_1,\Phi_2, \Phi_3\right)\nn\\
		&=& -\frac{1}{2} m_{11}^2  \left(\Phi_1^\dagger\Phi_1\right)  - \frac{1}{2} m_{22}^2  \left(\Phi_2^\dagger\Phi_2\right) + \frac{\lambda_1}{2} \left(\Phi_1^\dagger\Phi_1\right)^2 +\frac{\lambda_2}{2} \left(\Phi_2^\dagger\Phi_2\right)^2\nn\\
		&&  + \lambda_3 \left(\Phi_1^\dagger\Phi_1\right)  \left(\Phi_2^\dagger\Phi_2\right) + \lambda_4 \left(\Phi_1^\dagger\Phi_2\right) \left(\Phi_2^\dagger\Phi_1\right) + \frac{1}{2} \left[\lambda_5\  \left(\Phi_1^\dagger\Phi_2\right)^2 +\ h.c. \right]\nn
		\\
		&&  -\frac{1}{2} m_{33}^2\ \Phi_3^*\Phi_3 + \frac{\lambda_8}{2} \left( \Phi_3^*\Phi_3 \right)^2   + \lambda_{11} \left\vert \Phi_1\right\vert^2 \Phi_3^*\Phi_3 + \lambda_{13} \left\vert\Phi_2\right\vert^2 \Phi_3^*\Phi_3
		\nn\\
		&&  -i\, \kappa\,\, \left[\left( \Phi_1^\dagger\Phi_2 + \Phi_2^\dagger \Phi_1 \right)\,\,\left(\Phi_3-\Phi_3^\star\right)  \right]
		\label{eq:scalarpot}
		\eea
		where 
		\bea
		&&\Phi_1\equiv\left[\begin{array}{c} \phi_1^+\\
			\frac{1}{\sqrt{2}} \,\left(v_{\rm SM}+\phi_1^0 + i\,\eta_1^0\right)\\\end{array}\right];\,\, \, \Phi_2\equiv\left[\begin{array}{c} \phi_2^+\\
			\frac{1}{\sqrt{2}} \,\left( \phi_2^0 + i\,\eta_2^0\right)\\ \end{array}\right]\,\, \, {\rm and} \,\Phi_3\equiv \frac{1}{\sqrt{2}}\, \left(v_s +\phi_3^0 + i\,\eta_3^0 \right) \label{scalarfields}
		\eea
		and $D_\mu \Phi_i$($i=1,2,3$) is the covariant derivative for the field $\Phi_i$. 
	\end{widetext}
\end{subequations}
where all couplings in the scalar potential and Yukawa sector    are real in order to preserve the CP invariance. Here, we have invoked an additional global $U(1)$ symmetry under $\Phi_3 \to e^{ i\,\alpha} \Phi_3$  to reduce the number of free parameters in the scalar potential, which is however allowed to be softly broken by the $\kappa$ term.
Further, the Yukawa terms are given by
\begin{widetext}
		\bea
		-{\cal L}_{\rm Yukawa} &=& y_{u}\ \overline{Q_L}\ \widetilde{\Phi_1}\ u_R\ +\ y_{d}\ \overline{Q_L}\ \Phi_1\ d_R\ +\ y_{{}_{l}}\ \overline{l_L}\ \Phi_1 \ e_R\ + y_1\ \overline{l_L}\ \Phi_2 \ e_R  +\ \text{h.c.} \label{YukawaLag}
		\eea
		
\end{widetext}
\par The  stability of the scalar potential given in \eqref{eq:scalarpot} has been discussed in the article 
\cite{Bharadwaj:2021tgp}  and  the reader  may refer to  it  for the co-positivity conditions on the scalar potential and its minimisation. 

\par The absence of  mixing among the imaginary  component of the inert doublet with the real component of either the first SM like doublet or the singlet results in the decoupling of the mass matrices  for neutral scalars and pseudoscalars.
The $2\times 2$  CP-even  neutral scalar  mass matrix arises due to the mixing of the real components of SM like first doublet $\Phi_1$ and the singlet $\Phi_3$. 
Diagonalisation of this CP-even mass matrix by orthogonal rotation matrix parameterized in terms of the mixing angle $\theta_{13}$
results in two  mass eigenstates $h_1$ and $h_3$ with masses given by
\begin{subequations}
	\bea
	m_{h_1}^2&=& \cos^2 {\theta_{13}}  \,\,\lambda _1 \, \vsm^2+\sin  \left(2 {\theta_{13}}\right)  v_{s}\,\, \lambda _{11} \,\,\vsm \nn\\
	&& +\sin^2 {\theta_{13}} \,\, v_{s}^2\,\, \lambda _8 \label{eq:mh1}\\
	m_{h_3}^2&=& \sin ^2 {\theta_{13}}  \,\, \lambda _1\,\, \vsm^2-\sin  \left(2 {\theta_{13}} \right) v_{s} \,\,\lambda _{11} \,\, \vsm \nn\\ 
	&& +\cos^2 {\theta_{13}} \,\, v_{s}^2\,\, \lambda _8  \label{eq:mh3}\\
	\text{with} \qquad\qquad && \nn\\
	\tan 2\theta_{13}&=&\frac{\lambda_{11} \, \vsm v_{s}}{\l_1 \, \vsm^2-\l_8 v_{s}^2}  \label{eq:tantheta13}
	\eea
\end{subequations}
\noindent
Similarly, the diagonalisation of  mass matrix for CP-odd scalars $\eta_2^0$ and  $\eta_3^0$ 
gives the pseudoscalar mass eigenstates $A^0$ and $P^0$ with masses 
given by 
\begin{subequations}
	\bea
	m_{A^0}^2&=& \frac{1}{2} \left( \overline{\lambda}_{345}\,  \vsm^2- m_{22}^2 +\l_{13} v_{s}^2\right)\cos^2\theta_{23} \nn\\
	&& - \sqrt{2} \kappa \, \vsm \sin2\theta_{23} \label{eq:mA0}\\
	m_{P^0}^2&=& \frac{1}{2} \left( \overline{\lambda}_{345}\, \vsm^2- m_{22}^2 +\l_{13} v_{s}^2\right)\sin^2\theta_{23}\nn\\
	&& + \sqrt{2} \kappa \, \vsm \sin2\theta_{23}  \label{eq:mP0}
	\eea 
	where $\overline{\lambda}_{345}=\lambda_3+\lambda_4-\lambda_5$ and 
	the mixing angle $\theta_{23}$ is given by 
	\begin{eqnarray}
		\kappa = -\,\frac{1}{2\,\sqrt{2} \, \vsm  }\left(m^2_{P^0}+m^2_{A^0}\right)\tan\left(2\,\theta_{23}\right)
	\end{eqnarray}
\end{subequations}
Out of the remaining neutral and charged scalar mass eigenstates,  $\eta_1^0$ and $\phi_1^\pm$ are the massless Nambu-Golsdstone  bosons and the masses of $\phi_2^0$ and  $\phi_2^\pm$ which are renamed as $h_2$ and $ H^\pm$ respectively are given by 
\begin{subequations}
	\bea
	m_{h_2}^2&=&\frac{1}{2}\ \left[  -m_{22}^2+\  \left(\lambda _3 + \lambda _4 + \lambda _5\right)\, \vsm^2 + \l_{13} v_{s}^2\right] \nn\\
	&& 	\label{eq:mh2a} \\
	m_{H^\pm}^2&=&- m_{22}^2\ + \lambda_3 \vsm^2+\ \l_{13} v_{s}^2 \label{eq:mhpm}
	\eea
\end{subequations}
It should be noted that the parameter $\lambda_2$ appears only in the quartic interaction of $Z_2 -$
odd particles coming from the inert doublet $\Phi_2$ and does not contribute to the mass spectrum. It is therefore not constrained by our analysis.

The remaining eleven parameters in the scalar potential~\eqref{eq:scalarpot}, namely,
$m_{11}$, $ m_{22}$, $m_{33}$, $\lambda_{i= 1, 3, 4, 5, 8, 11,13}$  and $ \kappa$
can now be expressed in terms of the VEVs, masses and mixing angles:
\bea
\vsm ,\, v_s, \, m_{22}^2, \, m_{h_1}^2,\, m_{h_2}^2,\, m_{h_3}^2,\, m_{H^\pm}^2,\, m_{A^0}^2,\, m_{P^0}^2,\,  \theta_{13}, \, \theta_{23} \nn \\
\label{eq:parameters1}
\eea
For the relations among the mass parameters and scalar couplings of the Lagrangian, the reader is referred to the  appendix~\ref{app:coup-mass}.
 Further,  the  dimension-full scalar triple couplings of the charged Higgs bosons with neutral scalars are expressed as $g_{h_iH^+H^-}=(\vsm \, \lambda_{h_iH^+H^-} )$, where
\begin{subequations}
	\bea 
	\lambda_{h_1 H^+ H^-}  &=& \lambda_3 \,\cos\theta_{13} +  \frac{v_s}{\vsm} \,\lambda_{13}\,\sin\theta_{13} \label{eq:gh1hphm}\\
	\lambda_{h_3 H^+ H^-}  &=&   \frac{v_s}{\vsm} \lambda_{13}  \cos\theta_{13}\ -\ \lambda_3\  \sin\theta_{13}	 
	\label{eq:gh3hphm}
	\eea
\end{subequations}
are the dimensionless couplings.    

The Yukawa interactions given in \eqref{YukawaLag} 
can be rewritten in terms of mass eigenstates as 
\begin{widetext}
	\bea
	-{\cal L}^{^{\rm Yukawa}}_{\rm SM\, Fermions}&=&\sum_{s_i\equiv h_1,h_3}\frac{y_{_{ffs_i}}}{\sqrt{2}} \left(\vsm \,\,\delta_{s_i,h_1} + s_i\right) \bar f\ f\ + \frac{y_{_{llh_2}}}{\sqrt{2}} (h_2\ \bar l^- \ l^- ) + \sum_{s_i\equiv P^0,A^0} \frac{y_{_{lls_i}}}{\sqrt{2}} (s_i\ \bar l^-  \gamma_5\ l^-) \nn\\
	&& +\ \left[y_{_{l\nu H^-}}\  (\bar \nu_l\ P_R\ l^-   H^+ )+  {\rm h.c.} \right],
	\label{eq: L-yukawamasseigenstates}
	\eea
\end{widetext}
where $f$ and  $l^-$ represent SM fermions and SM charged leptons respectively. The Yukawa couplings  with scalar/ pseudoscalar mass eigenstates  are listed  in table \ref{Table:Yukawa}.  
\begin{table}[h]\footnotesize
	\begin{center}
		\begin{tabular}{cc||cc}
			\hline\hline 
			\rule{0mm}{5mm}
			$y_{_{ffh_1}}\,\,\,$ & $\quad \left(\sqrt{2} m_f/ v_{\rm SM}\right) \cos\theta_{13}$   &  $\,y_{_{llh_2}}\,\,\,$&$y_1$\\[0.2cm]
			$y_{_{ffh_3}}\,\,\,$&$- \left(\sqrt{2}m_f/ v_{\rm SM}\right)  \sin\theta_{13}$  & $\,y_{_{llP^0}}\,\,\,$&$-i\ y_1 \sin\theta_{23}$\\[0.2cm]
			$y_{_{l\nu H^-}}$ &$y_1$   &$\,y_{_{llA^0}}\,\,\,$&$\quad i\ y_1 \cos\theta_{23}$ 
			\\
			&&&\rule[1ex]{0pt}{0mm} \\
			\hline\hline
		\end{tabular}
	\end{center}
	\caption{\em{Yukawa couplings}}
	\label{Table:Yukawa}
\end{table}
\section{Constraints on Parameter Space}
\label{sec:constraints}
The  theoretical constraints and existing experimental
observations  restrict the parameter space of  any model beyond the SM. The following
physical parameters of the  model  affect the observables considered by us in this article:
\bea
{\text{Masses}}&:& m_{h_1},\, m_{h_2},\, m_{h_3},\, m_{H^\pm},\,\ m_{A^0},\, m_{P^0} \nn\\
{\text{Mixing Angles}}&:& \theta_{13},\, \theta_{23} \nn\\
{\text{Couplings}}&:&  y_1, \, \lambda_{{h_1 H^+ H^-}},\, \lambda_{{h_3 H^+ H^-}}
\label{eq:parameters2}
\eea
We discuss below various constraints imposed on these parameters.
\subsection{Theoretical Constraints}
Let us first consider theoretical limitations on the scalar potential of our Model. 
The scalar potential given in~\eqref{eq:scalarpot} should satisfy the stability and co-positivity conditions listed in reference~\cite{Bharadwaj:2021tgp}. Further,
tree level perturbative unitarity requires  that
\be 
\left\vert\lambda_i\right\vert \le 4 \pi, \quad \text{and}\quad \left\vert y_1\right\vert < \sqrt{4\pi}.
\label{eq:perturbative}
\ee
where $\lambda_i$ are all the quartic scalar couplings and $y_1$ is the Yukawa coupling.
\par The  relations  among mass parameters and scalar couplings of the Lagrangian,  along with the co-positivity conditions result in the following  two mutually exclusive allowed regions of parameter space:
\begin{widetext}
	\be
	\Theta(\left\vert \lambda_5\right\vert-\lambda_4)=
	\begin{cases} \Theta\left[m_{H^\pm}^2-(m_{A^0}^2+m_{P^0}^2)\right] &\,\,\text{for}\,\,\,\, m_{h_2}^2>m_{A^0}^2+m_{P^0}^2 \, : \quad  \text{Region I} \\
		\Theta\left[m_{h_2}^2-m_{H^\pm}^2\right]
		& \,\,\text{for}\,\,\,\, m_{h_2}^2<m_{A^0}^2+m_{P^0}^2  \, : \quad \text{Region II} 
	\end{cases}
	\label{eq:derivedpositivity}
	\ee
\end{widetext}
These two regions I and II correspond to $\l_5>0$ and
	 $\l_5<0$ respctively (as per  equation~\eqref{eq:lam5} in appendix~\ref{app:coup-mass}) 
In this article we explore the  phenomenology rich  region I given by
\be
m_{h_2}^2>m_{A^0}^2+m_{P^0}^2  \quad \text{and} \quad
m_{H^\pm}^2>m_{A^0}^2+m_{P^0}^2 . 
\label{eq:lowlim-mh2mhp}
\ee
Given the aforementioned mass hierarchy, no viable scalar dark matter exists in this region. Also, the non-vanishing Yukawa coupling $y_1$ in the Lagrangian \eqref{YukawaLag} prevents the lightest pseudoscalar from being a dark matter candidate by permitting the pseudoscalar to decay to leptons.

Now we consider the constraints from some experimental observations in the next section. In all these calculations, the values  of parameters $\alpha$, the Fermi constant $G_F$ and $Z$ boson mass $\mz$ are taken to be the measured values~\cite{Workman:2022ynf}. 
\subsection{Constraints from Higgs Decay}
\label{subsec:H-decay}
Since, LHC data favors a scalar eigenstate $H$ with mass$\,\sim 125\,$GeV~\cite{Workman:2022ynf}, we  identify CP even lightest neutral scalar $h_1$,coming predominantly from the doublet
$\Phi_1$(equation\eqref{eq:mh1}) with the observed scalar $H$ and take $m_{h_1} = 125\,$GeV. Further, the  couplings of  $h_1$ with a pair of fermions and gauge bosons are the corresponding SM Higgs couplings but suppressed by  $\cos\theta_{13}$ due to $\Phi_1-\Phi_3$ mixing. 

We now compare the total Higgs decay width in SM~ \cite{Denner:2011mq, LHCHiggsCrossSectionWorkingGroup:2013rie}
\be 
\Gamma(h^{\rm SM}\to \textrm{all})\sim 4.07 \, \text{MeV}
\label{eq:gam_hsmtoall}
\ee 
with the recently measured total Higgs decay width at the Large Hadron Collider(LHC)~\cite{Workman:2022ynf} 
\be
\Gamma(H \to \text{all})_{\rm LHC}  = 3.2^{+2.4}_{-1.7}\, \text{MeV} .
\label{eq:gam_Htoall}
\ee

\par
We examine the bounds on  partial decay widths of  125 GeV $h_1$ at LHC and determine the constrained parameter space by demanding that, in our model, $h_1$ decays can account for the measured value of the total Higgs decay width. To this end,  we define the signal strength  $\mu_{{}_{XY}}$ {\it w.r.t.} $h_1$ production {\it via} dominant gluon fusion in $p-p$ collision, followed by its  decay to  $X \, Y$ pairs in the narrow width approximation as
\begin{widetext}
	\bea
	\mu_{{}_{XY}}&=&\frac{\sigma(pp\to h_1\to XY)}{\sigma(pp\to h\to XY)^{\textrm  {SM}}}=\frac{\Gamma\left( h_1\to g\, g\right)}{\Gamma \left(h^{SM} \to g\, g\right)}\,\,\,\, \frac{\textrm{BR}\left(h_1\to X\,Y\right)}{\textrm{BR}\left(h^{\rm SM}\to X\,Y\right)}
	=\cos^2\theta_{13}\,\,\frac{\textrm{BR}\left(h_1\to X\,Y\right)}{\textrm{BR}\left(h^{\rm SM}\to X\, Y\right)}
	\eea
\end{widetext}
\onecolumngrid
\begin{center}
	\begin{figure}[htb]
		\subfloat[\justifying \em{The solid red curve shows the variation of  $\mu_{{}_{W W^\star}}$ computed in our model with the CP-even mixing angle $\theta_{13}$. The shaded blue region depicts the  allowed one sigma region for the measured $\mu_{{}_{W W^\star}} = 1.00 \pm 0.08$ \cite{Workman:2022ynf}.}\label{fig:Hdecay1} ]{
			\includegraphics[width=.48\linewidth]{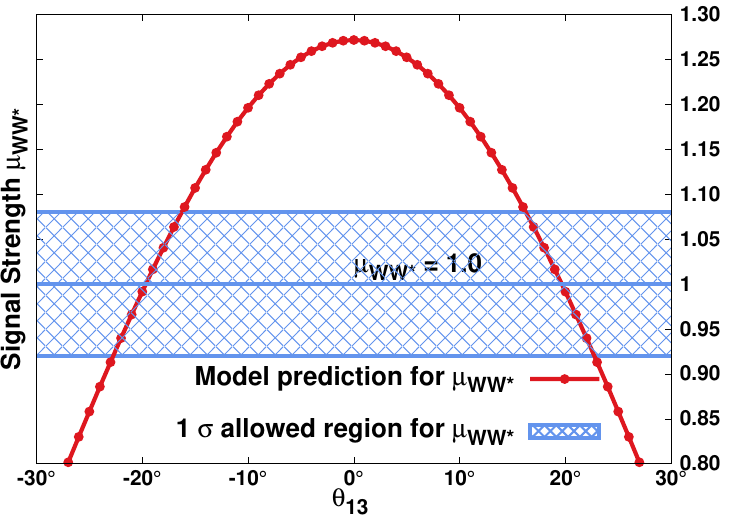}
		}\hfill%
		\subfloat[\justifying \em{ Color density map for the constraint $\mu_{{}_{\gamma \gamma}}/\mu_{{}_{W W^\star}}$ $= 1.1\pm0.11$ at $2\sigma$ level~\cite{Workman:2022ynf} corresponding to the $\theta_{13} = 20^\circ$ in  $  \lambda_{{h_1 H^+ H^-}} - \mhpm$ plane.}\label{fig:Hdecay2}]{%
			\includegraphics[width=.48\linewidth]{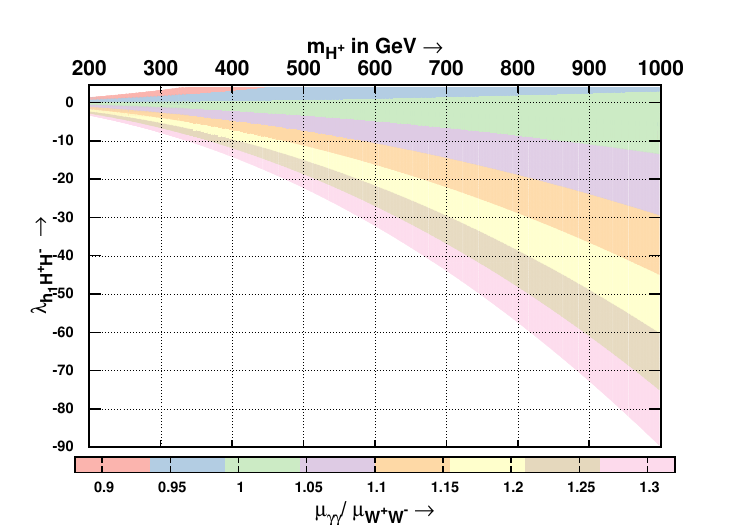}}%
		\caption[nooneline]{\justifying \em{Constraints on parameters $\theta_{13}$,  $ g_{_{h_1 H^+ H^-}}$ and  $ \mhpm$ from the measurements of the  partial Higgs decay widths to the gauge bosons at the LHC \cite{Workman:2022ynf}. }}
	\end{figure}
\end{center}
\twocolumngrid
%
\par The partial decay width of $h_1\to W\,W^\star$ channel is related to the corresponding value in SM as
\bea
\Gamma(h_1\to W W^\star)&=& \cos^2 \theta_{13}\ \Gamma(h^{\rm SM} \to W W^\star)
\label{h1toww}
\eea
giving the signal strength 
\bea
\mu_{{}_{WW^\star}}&=& \cos^4 \theta_{13}\ \frac{\Gamma(h^{\rm SM} \to \text{all})}{
	\Gamma(H \to \text{all})_{\rm LHC} } \simeq 1.27   \cos^4 \theta_{13} 
\nn\\
\label{eq:wwsignal}
\eea
Thus, the signal strength $\mu_{{}_{WW^\star}}$ depends only one parameter of the model, namely $\theta_{13}$ which  can be strongly constrained by the observed value,  $\mu_{W W^\star} =  1.00 \pm 0.08$ \cite{Workman:2022ynf}.  The one sigma band around the central value of the observed $\mu_{W W^\star} $ is shown in the  figure~\ref{fig:Hdecay1},  which restricts  the  value of $\theta_{13}$  to 
$ 19.7^\circ  \lesssim \left\vert\theta_{13}\right\vert \lesssim 22.8^\circ$. 
Throughout this work, we take $\theta_{13} = 20^\circ$.

We now calculate the partial  decay width of  $h_1 \to \gamma \gamma$ channel
at one loop  in our  model that may be  parameterized as
\begin{eqnarray}
	\Gamma(h_1\to\gamma\,\gamma) &=& \cos^2\theta_{13} \left\vert 1 + \zeta_{\gamma\gamma}\right\vert^2\,\Gamma\left( h^{\rm SM} \to \gamma\,\gamma\right) 
	\label{h1toaa},
\end{eqnarray}
where the SM Higgs partial decay width in $\gamma\,\gamma$  channel and the dimensionless parameter $\zeta_{\gamma\gamma}$ 
are given by~\cite{Bonilla:2014xba, Bharadwaj:2021tgp}
\begin{widetext}
	\begin{subequations}
		\bea
		\Gamma(h^{\rm SM}\to\gamma\gamma)&=&\frac{G_F\alpha^2\ m_{h}^3}{128\sqrt{2}\pi^3}\ \left\vert  \frac{4}{3}  {\cal  M}^{\gamma\gamma}_{1/2}\bigg(\frac{4m_t^2}{m_{h}^2}\bigg)+  {\cal  M}^{\gamma\gamma}_1\bigg(\frac{4m_{\rm W}^2}{m_{h}^2}\bigg) \right\vert^2\label{hsmtoaa}\\
		\zeta_{\gamma\gamma} &=& \frac{v_{\rm SM}}{\cos\theta_{13}}\left[\frac{\frac{g_{h_1H^+H^-}}{2\,m^2_{H^\pm}} {\cal M}^{\gamma\gamma}_0\left(\frac{4 m_{H^\pm}^2}{ m_{h_1}^2} \right)  }{{\cal M}^{\gamma\gamma}_1\left(\frac{4\,m_{\rm W}^2} {m_{h_1}^2}\right) + \frac{4}{3}\, {\cal M}^{\gamma\gamma}_{1/2}\left(\frac{4\,m_t^2}{m_{h_1}^2}\right)}\right]
		\label{zetaaa}
		\eea
		The loop form factors ${\cal M}^{\gamma\gamma}_{0,\,1/2,\,1}$  in the above equations  are defined in the appendix \ref{app:formfactors}.
	\end{subequations}
	Using the relations \eqref{h1toww} and \eqref{h1toaa},  the ratio of signal strengths  becomes
	%
		\bea
		\frac{\mu_{{}_{\gamma \gamma}}}{\mu_{{}_{W \,W^\star}}}&=& \frac{\Gamma(h_1 \to \gamma \,\gamma)}{\Gamma(h_1 \to W\, W^\star)} \times \frac{\Gamma(h^{\rm SM} \to W W^\star)}{\Gamma(h^{\rm SM} \to \gamma \,\gamma)} \, = \,   \left\vert 1 + \zeta_{\gamma\gamma} \right\vert^2
		\label{aabywratio}
		\eea
\end{widetext}
 The average experimental values of signal strengths   $\mu_{{}_{\gamma \gamma}}=1.10\pm{+0.07}$ and $\mu_{{}_{W W^\star}}=1.00 \pm 0.08$~\cite{Workman:2022ynf} give $\mu_{{}_{\gamma \gamma}}/\mu_{{}_{W W^\star}}=1.1\pm 0.11$. The  value for this ratio in our model depends only upon the parameters $\theta_{13}$, $\mhpm$ and  $ \lambda_{{h_1 H^+ H^-}}$. Varying the $m_{H^+}$ between 210 GeV - 1 TeV and fixing $\theta_{13}=20^\circ$, we find that the one sigma and two sigma constraints on $\mu_{{}_{\gamma \gamma}}/\mu_{{}_{W W^\star}}$ restricts the charged Higgs couplings to the lightest CP even scalar within a allowed range. This allowed range depends upon the value of \mhpm. For example,  
 	for $ \mhpm = 1\, $TeV, the  range  allowed  by $\mu_{{}_{\gamma \gamma}}/\mu_{{}_{W W^\star}}$ is 
\bea
-60 <\lambda_{{h_1 H^+ H^-}}< 3 &&  {\rm at} \quad  1\sigma\nonumber\\
-90 <\lambda_{{h_1 H^+ H^-}}< 4 && {\rm at} \quad  2\sigma
\eea 
In the figure~\ref{fig:Hdecay2}, we exhibit the contours satisfying $\mu_{{}_{\gamma \gamma}}/\mu_{{}_{W W^\star}}$ at $2\sigma$ level for $\theta_{13} = 20^\circ$  in  the $ \lambda_{{h_1 H^+ H^-}} - \mhpm$ plane. 

\par Since the experimental uncertainty for $\mu_{{}_{Z \gamma }}$ \cite{Workman:2022ynf} is large, we do not expect any more constraints on the  the model parameters from $h_1\to Z\,\gamma$ decay channel~\cite{Bharadwaj:2021tgp}. 
\subsection{Constraints from LEP II Data}
\label{subsec:LEP}
The scalar and pseudoscalar masses along with  the Yukawa coupling $y_1$  in our  model can be constrained from the existing LEP II data either by investigating the (a) direct pair production of scalars and pseudoscalars  or  (b) by production of pair of fermions mediated by these additional  physical scalars or pseudoscalars.  The direct neutral scalar and pseudoscalar pair production channels 
\bea
e^+\ e^- \to Z^\star \to A^0 / P^0 + h_i
\eea
constraint the  sum of neutral Higgs masses ($\sum_{i=1}^3 m_{h_i} + m_{A_0} + m_{P^0}$) to be $\gtrsim$ 200 GeV ~\cite{ALEPH:2013dgf}.  To be consistent with these bounds from LEP, we perform our analysis for all scalar and pseudoscalar mases above 210 GeV. 

\par The  production cross section of  fermion pairs gets a contribution from additional scalars and pseudoscalars in the model through  new leptonic Yukawa coupling $y_1$. This additional contribution should be in agreement with the  electroweak precision measurements conducted by LEP experiments. The combined analysis of DELPHI and L3 at LEP II at
$\sqrt{s}=200\ GeV$  estimate the  cross-section of muon pair production to~\cite{ALEPH:2013dgf}  
\be 
\sigma (e^+\ e^- \to \mu^+\ \mu^-)=3.072 \pm 0.108 \pm 0.018\, { \rm pb } .
\label{eq:mu-cross-LEP}
\ee
The  excess contribution to  this cross section in our model over the  SM one  can be written as
\begin{widetext}
	\bea
	\sigma_{\mu^+ \mu^-}^{\rm Excess}&=&\frac{s}{64 \pi}\ \sqrt{\frac{s-4 m_\mu^2}{s-4m_e^2 } }\, \times \nn\\
 &&  \left[ y_1^2  \left( -\frac{\cos^2\theta_{23}}{s-m_{A^0}^2}  - \frac{\sin^2\theta_{23}}{s-m_{P^0}^2} 
	+  \frac{1}{s-m_{h_2}^2} \right) + \frac{2m_e m_\mu}{\vsm^2 } \,  \left( \frac{\cos^2\theta_{13}}{s-m_{h_1}^2}  + \frac{\sin^2\theta_{13}}{s-m_{h_3}^2}  \right)  \right]^2 \nn\\
	&&
	- \left[\frac{2m_e m_\mu}{\vsm^2 } \,  \left( \frac{1}{s-m_{h_{_{SM}} }^2} \right) \right]^2
	\label{xsec:mumu}
	\eea
\end{widetext}
\par
\begin{center}
	\begin{figure}[htb]
		\subfloat[{ \em{ $\!\! m_{_{A^0}}\!\! \! =\!600$GeV, $ R_P \! =\! 0.5$ }}\label{fig:LEP3a} ]{
			\includegraphics[width=.32\linewidth]{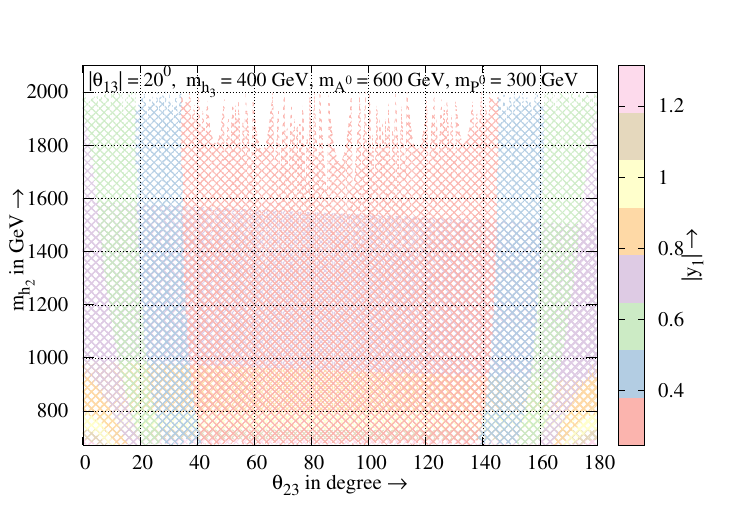}
		}\hfill
		\subfloat[{ \em{$\!\! m_{_{A^0}}\!\! \! =\!300$GeV, $ R_P \! =\! 1$}}\label{fig:LEP3b} ]{%
			\includegraphics[width=.32\linewidth]{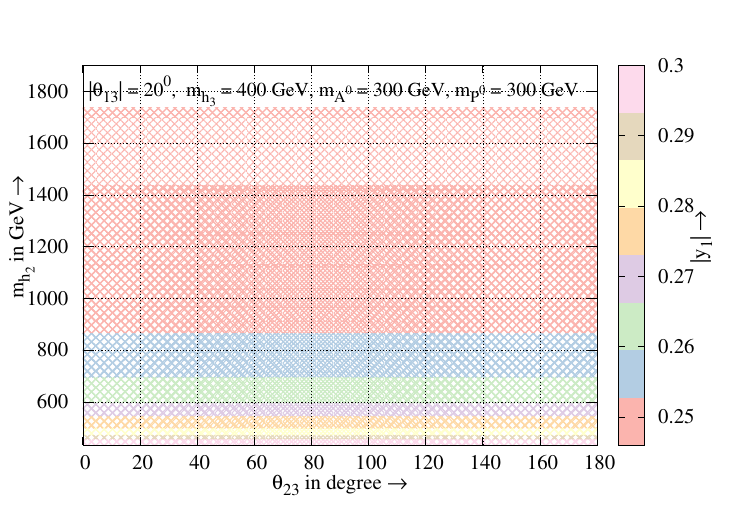}
		}\hfill
		\subfloat[{ \em{$\!\! m_{_{A^0}}\!\! \! =\!300$GeV, $ R_P \! =\! 2$}}\label{fig:LEP3c}] {%
			\includegraphics[width=.32\linewidth]{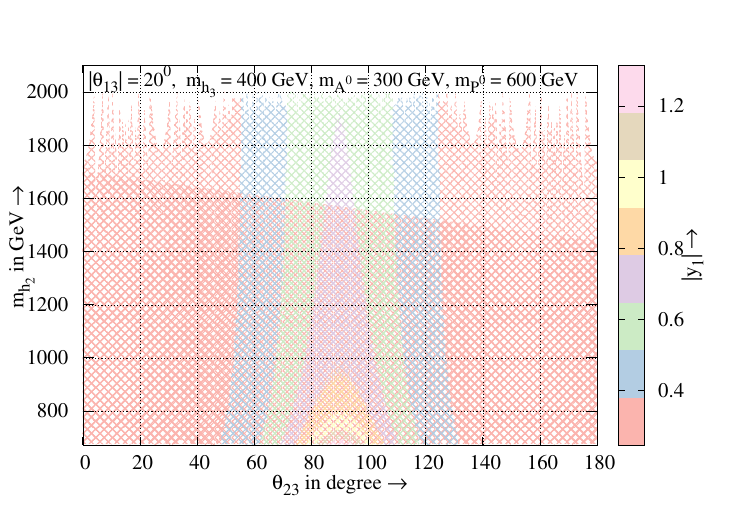}%
		}\hfill
		\caption[nooneline]{\justifying \em{Yukawa  coupling $\left\vert y_1\right\vert $ color density maps for $\theta_{13} =20^\circ$  and $m_{_{h_3}} $ = 400 GeV  in the  $m_{_{h_2}}- \theta_{23}$ plane  satisfying the constraints from combined analysis of DELPHI and L3, namely, $\sigma (e^+\ e^- \to \mu^+\ \mu^-)=3.072 \pm 0.108 \pm 0.018$ pb at $\sqrt{s}=200\ GeV$  \cite{ALEPH:2013dgf}  corresponding to the  three  parameter sets (a) $m_{_{A^0}} = 600$GeV, $ R_P  (= m_{_{P^0}}/m_{_{A^0}}) =  0.5$, (b) $m_{_{A^0}} = 300$GeV, $ R_P  =1$, (c)  $m_{_{A^0}} = 300$GeV, $ R_P = 2$.  }  }
		\label{fig:LEP3}
	\end{figure}
\end{center}

 We compute this contribution to  $\mu$-pair production cross-section  given by  
equation~\eqref{xsec:mumu} and put  constraints on the model parameters by accommodating 
this excess contribution within the 1$\sigma$ uncertainty (0.1095pb)  in the  cross-section 
$\sigma (e^+\ e^- \to \mu^+\ \mu^-)$  given by~\eqref{eq:mu-cross-LEP}. The figure~\ref{fig:LEP3}  depicts the  density maps for  $\left\vert y_1\right\vert $ in the  $m_{_{h_2}}- \theta_{23}$ plane corresponding to  various values of  $m_{_{A^0}}$ and pseudo scalar mass ratio  $R_P = m_{_{P^0}}/m_{_{A^0}}= 0.5, \, 1, \, 2$. The value of  $m_{_{h_3}} $ is taken to be 400~GeV in these plots.  For a given $m_{_{A^0}}$ and $m_{_{P^0}}$, the value of $m_{h_2}$ has a lower limit determined by~\eqref{eq:lowlim-mh2mhp}.

Following  observations may be noted from  the equation~\eqref{xsec:mumu}:
\ben
\item The cross-section is found to be less sensitive to the variation of $m_{h_3}$ since, for $\theta_{13}\approx 20^\circ$, the term proportional to $\left(m_e m_\mu\right) / \vsm^2 $ is negligibly tiny. This enables the  LEP data to tightly constrain the magnitude of the $\left\vert y_1\right\vert$ and $\left\vert\theta_{23}\right\vert$ for the varying scalar and pseudo-scalar masses upto a TeV scale.
\item The permitted  range  of $\left\vert y_1\right\vert$  is primarily governed by the choice of $\theta_{23}$ and  the pseudoscalar mass ratio  $R_P = m_{_{P^0}}/m_{_{A^0}}$. With the exception of $R_P=1$, we note that  the allowed values of $\left\vert y_1\right\vert $ are not very sensitive to  $m_{h_2}$. This is because,   the matrix element squared $ \left\vert  {\cal M}_{\mu^+ \mu^-}^{\rm NP}\right\vert^2 $ in equation~\eqref{xsec:mumu} becomes independent of $\theta_{23}$ for $m_{A^0}=m_{P^0}$, and hence  the allowed values of $\left\vert y_1\right\vert$ are dictated by vlaues of $m_{h_2}$ and $m_{_{A^0}}$. This is evident from the $\left\vert y_1\right\vert $ color density map given in figure~\ref{fig:LEP3b}.
\item The color density maps in Figures~\ref{fig:LEP3a} and \ref{fig:LEP3c} show the concave and convex profiles of $\left\vert y_1\right\vert$  {\it w.r.t.} $\theta_{23}$ for $R_P<1$ and $R_P>1$, respectively, due to the presence of $\cos^2(\theta_{23})$ and $\sin^2(\theta_{23})$ with the respective  propagators for pseudoscalars $A^0$ and $P^0$. The  convexity/ concavity profile is more pronounced for lower scalar and pseudoscalar masses.
\een
\onecolumngrid
\twocolumngrid

\par The dominant direct charged Higgs pair production channels at the $e^+e^-$ collider:
\bea
e^+\ e^- &\to& \gamma^\star/\, Z^\star \to H^+ + H^-
\eea
limits the charged Higgs mass  between $(80-100)$ GeV \cite{ALEPH:2013htx}.  Assuming the branching ratio for the model predicted dominant decay channel of charged Higgs $\text{Br}(H^+\to \tau + \nu_\tau)$ to be unity, the ALEPH collaboration at LEP \cite{ALEPH:2013htx} gives the combined 95\% C.L. lower bound of 94 GeV on the mass of the charged Higgs boson. The LEP constraints on the masses of pseudoscalars and  the model restriction $\Theta\left[m_{H^\pm}^2-(m_{A^0}^2+m_{P^0}^2)\right] $ as given  in equation \eqref{eq:lowlim-mh2mhp} ensures that the probed charged Higgs mass is significantly higher than the lower bound obtained from LEP.
\par We now proceed  to look for viable regions of the parameter space already constrained in this section that accounts for the observed  measurements of the anomalous  magnetic dipole moment  for muon and the $W$-boson mass in the next two sections.
	\onecolumngrid
\begin{center}
	%
	\begin{figure}[htb]
		\subfloat[{ \em  Leptons  1-Loop}\label{fig:leptoneloop} ]{
			\includegraphics[height=2.6cm,width=.18\linewidth]{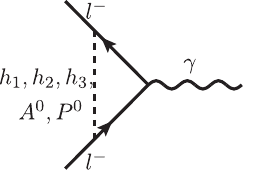}
		}
		\subfloat[{ \em{ Charged Scalars  1-Loop}}\label{fig:ChargedscalOneloop}]{%
			\includegraphics[width=.17\linewidth]{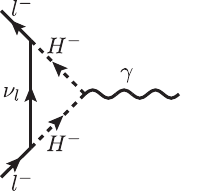}}\hspace{1.8em  plus 1fill}
		\subfloat[{ \em BarrZee  top Triangle}\label{fig:HPM-top} ]{
			\includegraphics[width=.17\linewidth]{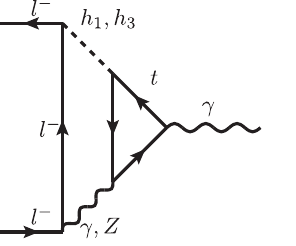}
		}
		\subfloat[{ \em BarrZee $H^\pm$  Triangle}\label{fig:HPM-triangle} ]{
			\includegraphics[width=.17\linewidth]{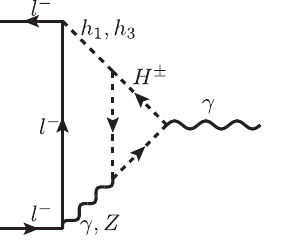}}
		\subfloat[{ \em{ BarrZee $H^\pm$ Bubble }}\label{fig:HPM-bubble}]{%
			\includegraphics[width=.17\linewidth]{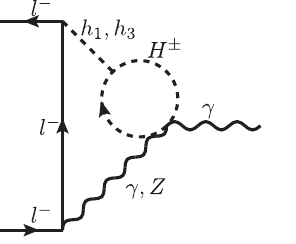}}%
		\caption[nooneline]{\justifying \em{One-loop and two-loop dominant diagrams contributing to  $g-2$ of charged lepton $l$.}}
		\label{fig:FD-MDM}
	\end{figure}
\end{center}
\twocolumngrid
\section{Calculation of  $\mathbf{(g-2)_\mu}$ and $\mathbf{W}$-boson mass} 
\label{sec:mdm-mw}
We explore in this and the following section, how our model can account for the positive pull in the observed  muon anomalous magnetic moment and the $W$-boson mass. We discuss the formalism for computing  both quantities in this section, while the next section provides  multivariate numerical analysis. 
\subsection{Muon Anomalous Magnetic Moment}
\label{subsec:mdm}
Now, we compute the dominant one- and two-loop contributions to the anomalous magnetic moment of a charged lepton $(l)$ in our model and then subtract the SM contributions from the same. This difference in the anomalous magnetic moment $\Delta a_l$ arises due to the exchange of the additional spectrum of charged and neutral scalars and pseudoscalars in the I2HDM at the one- and two-loop level of the perturbative calculations. Based on the Lagrangian given in equations \eqref{eq:scalarpot} and  \eqref{YukawaLag}, the dominant Feynman diagrams  at one loop and two loop Barr-Zee diagrams are given in the figure~ \ref{fig:FD-MDM}.  Note that the Barr-Zee diagrams involving W-bosons are not allowed by $Z_2$ symmetry.

The excess contribution to lepton $\Delta a_l$ at the one-loop level is given by
	\begin{widetext}
		\bea
		\delta a_l^{\rm 1\,loop }&=& \frac{1}{16\, \pi^2}\left[ 2\ \frac{m_l^4}{\vsm^2} \ \left( \frac{\cos^2\theta_{13}}{m_{h_1}^2}\  + \frac{\sin^2\theta_{13}}{m_{h_3}^2}- \frac{1}{m_{h^{\rm SM}}^2} \right)\ {\cal I}_1 +  m_l^2 \ \left( \frac{\cos^2\theta_{23}}{m_{A^0}^2} + \frac{\sin^2\theta_{23}}{m_{P^0}^2} \right)\  y_1^2\  {\cal I}_2 \right. \nn\\
		&&  \left.  + \frac{m_l^2}{m_{h_2}^2}\ y_1^2\ {\cal I}_1 + \left\vert y_1\right\vert^2\ \frac{m_l^2}{m_{H^\pm}^2}\ {\cal I}_3 \right]
		\label{eq:MDMoneloop}
		\eea
	\end{widetext}%
	\noindent 
	where the one loop integral functions ${\cal I}_1, \,{\cal I}_2$ and ${\cal I}_3$ are defined in the appendix \ref{app:MDMLoopFunc} in equations \eqref{i1eq},  \eqref{i2eq} and  \eqref{i3eq}, respectively. 
	We observe that the one-loop amplitudes in Figure \ref{fig:leptoneloop} are  negative and positive, corresponding to mediating pseudoscalars and  scalars, respectively, while  the contribution from the charged Higgs loop in Figure \ref{fig:ChargedscalOneloop} is negative  and competitively much smaller in magnitude. It is to be noted that for $m_{A^0} = m_{P^0}$, the one-loop contribution becomes independent of the mixing angle $\theta_{23}$. 
	
	\par The contributions of two loop diagrams, some of which may dominate inspite of an additional loop suppression factor play a crucial role in the estimation of anomalous MDM. It is shown in the literature that the dominant two-loop Barr-Zee diagrams mediated by neutral scalars and pseudoscalars can become relevant for certain mass scales so that their contribution to the muon anomalous MDM  are of the same order to that of one loop diagrams \cite{Chun:2020uzw}. The additional contributions to the lepton $\Delta a_l$  at   two-loop level  is given by 
	\begin{widetext}
		\begin{eqnarray}
			\delta {a_l}^{2\ loop} &=&  \frac{\alpha_{\rm em}}{4\ \pi^3}\ \left[ 
			\frac{m_l}{\vsm } \frac{m_t}{\vsm}\ \left\{\sin^2\theta_{13}\ f\left( \frac{m_t^2}{m_{h_3}^2}\right) - \cos^2\theta_{13}\ f\left(\frac{m_t^2}{m_{h_1}^2}\right) +  f\left(\frac{m_t^2}{m_{h^{\rm SM}}^2}\right) \right\} \right. \nn\\
			&& \left. -\frac{ m_l^2}{4} \, \frac{m_l}{\vsm^2} \left\{ \frac{ \cos\theta_{13}}{ m_{h_1}^2}\ g_{_{h_1 H^+ H^-}}\ \tilde{f}\left( \frac{m_{H^\pm}^2}{m_{h_1}^2} \right) - \frac{ \sin\theta_{13}}{ m_{h_3}^2}\ g_{_{h_3 H^+ H^-}}\ \tilde{f}\left( \frac{m_{H^\pm}^2}{m_{h_3}^2} \right) \right\} \right]
			\label{eq:MDM-2loop}
		\end{eqnarray}
	\end{widetext}
where the two loop integral functions $f$ and $\tilde f$ are given by  the equations \eqref{2loopint1} and \eqref{2loopint3} respectively in appendix \ref{app:MDMLoopFunc}.

\par It should be noted that the Barr-Zee diagrams of the type shown in figure~\eqref{fig:HPM-triangle}  with $W$ boson and charged Higgs $H^\pm$  replacing scalars ($h_1/h_3$) and
$W$ boson replacing $\gamma/Z$, that are usually present in a THDM do not contribute in this model because such a diagram will involve $WWh_2$ coupling which are forbidden in our model by the imposed $Z_2$ symmetry. 

With $m_{h_1} \approx m_{h^{\rm SM}} = 125\,$GeV and using the dimensionless couplings defined in equations~\eqref{eq:gh1hphm} and \eqref{eq:gh3hphm}, the two-loop Bar-Zee contribution in  equation\eqref{eq:MDM-2loop}  can be simplified to 
\begin{widetext}
	\begin{eqnarray}
		\delta {a_l}^{2\ loop} &=&  \frac{\alpha_{\rm em}}{4\ \pi^3}\ \left[ 
		\frac{m_l}{\vsm } \frac{m_t}{\vsm}\ \sin^2\theta_{13} \left\{ f\left( \frac{m_t^2}{m_{h_3}^2}\right) -  f\left(\frac{m_t^2}{m_{h_1}^2}\right) \right\} \right. \nn\\
		&-& \left. \frac{ 1}{4} \, \frac{m_l}{\vsm} \left\{ ( \l_{_{h_1 \!H^+ \!H^-} })\
		\frac{ m_l^2}{ m_{h_1}^2}\ \cos\theta_{13} \, \tilde{f}\left( \frac{m_{H^\pm}^2}{m_{h_1}^2} \right) - \l_{_{h_3 \! H^+ \! H^-}}\  \frac{  m_l^2}{ m_{h_3}^2}\ \sin\theta_{13}\, \tilde{f}\left( \frac{m_{H^\pm}^2}{m_{h_3}^2} \right) \right\} \right]
		\label{eq:MDM-2loop}
	\end{eqnarray}
\end{widetext}
Keeping the dimensionless parameers $\l_{_{h_i \! H^+ \! H^-}}$ 
($i=1,3$) reasonable value, say $\lesssim 10$. and with the range 
of masses considered by us (i.e. $m_{h_3}, m_{A^0},m_{P^0} 
>200 \,$GeV and $\mhpm > \sqrt{m_{A^0}^2 + m_{P^0}^2  } $), 
the supression due to factor $  m_l^2/ m_{h_i}^2$ in the second 
term (lower line in above equation) compared to 	$(m_l 
m_t/\vsm^2)$ makes the contribution of  charged Higgs to 
two-loop muon magnetic moment in figure \ref{fig:HPM-triangle} 
and \ref{fig:HPM-bubble}  negligibly small even for  $\l_{_{h_i \! 
		H^+ \! H^-}} $ as large as $ \sim 10^3$ .  However, the dimensionless 
	parameer  $\lambda_{{h_1 H^+ H^-}}$ is restricted from the 
	observation of signal strength ratio $\mu_{{}_{\gamma \gamma}}/\mu_{{}_{W W^\star}}$  as discussed in section~\ref{subsec:H-decay}
	while $\lambda_{{h_3 H^+ H^-}}$ is not restricted. 
\par  The Barr-Zee contributions are thus found to  dominantly depend on the mixing angle  $\theta_{13}$ and the scalar masses $m_{h_1}$ and $m_{h_3}$.

\par On analysing the combined contribution from the one- and two-loop diagrams for the constrained parameter space obtained in the preceding section, we find that, depending on the mass range of the scalars and pseudoscalars, both the one- and two-loop Barr-Zee contributions can be significant. In fact, by demanding the total anomalous magnetic dipole moment, to agree within one sigma of the measured value  as stated in equation~\eqref{eq:delamu}, we can further limit the parameter space. 
\subsection{$W$ Mass Computation}
\label{subsec:W-mass}
In this subsection, we compute the $W$-boson mass in our extended inert  two Higgs doublet models. The mass of the $W$-boson can be predicted from muon decay in terms of three  precisely measured quantities, namely, the Fermi constant, $G_\mu$, the fine structure constant, $\alpha_{\rm em}$, and the mass of the $Z$-boson, $\mz$ via
\bea
\mw^2 \left(1-\frac{\mw^2}{\mz^2}\right)  =  \frac{\pi \alpha_{\rm em} }{\sqrt{2} G_\mu   \left( 1 - \Delta r\right)}
\label{eq:wmass}
\eea
where $ \Delta r$ represents the quantum corrections to the relation and is a function of the scalar and pseudoscalar masses and the gauge couplings. This relationship is usually employed for predicting the
$W$-boson mass $\mw$ by an iterative procedure since $\Delta r$  is itself a function of $\mw$. The SM contribution to $ \Delta r$ at the full two-loop level, augmented by all the known three-loop contributions and the four-loop strong corrections, has been computed~\cite{Degrassi:2014sxa}. The discrepancy  between the measured and the SM value may be resolved via quantum corrections that modify $ \Delta r$.
Defining
$ (\Delta r)^\prime =   \Delta r \big\vert_{\rm NP} -   \Delta r \big\vert_{\rm SM} $
and using  measured values of $G_\mu$,  $\alpha_{\rm em}$ and  $\mz$ as input to the $SU(2) \times U(1)$ gauge theory, the relation
\bea
\mw^2 &= & (\mwsm)^2 \left( 1+ \frac{s_w^2}{c_w^2 - s_w^2}  
( \Delta r^\prime)   \right)
\label{eq:w-mass1}
\eea
gives the prediction of $W$-boson mass~\cite{Grimus:2008nb}.
Here $s_w = \sin{\theta_w}$ and $c_w = \cos{\theta_w}$, $\theta_w$ being the
weak angle and $(\Delta r^\prime)  $ represents the  measure of deviations of the quantum corrections in a new physics model from those in SM. It is possible to parameterize  $(\Delta r^\prime)  $  in terms of the oblique parameters  $S$, $T$ and $U$ as 
\bea
\Delta r^\prime  &= &  \frac{\alpha}{s_w^2} \left( - \frac{1}{2}\Delta S + c_w^2 \Delta T
+ \frac{c_w^2 - s_w^2}{4s_w^2} \Delta U \right).
\label{eq:delta-rp}
\eea
where $\Delta S$, $\Delta T$, and $\Delta U $ are the deviations from their corresponding SM values in the estimation of the oblique parameters in any new physics models \cite{Peskin:1991sw, *PhysRevLett.65.2963,  *Kennedy:1990ib, *Kennedy:1991sn, *Kennedy:1991wa, *Ellis:1992zi}. These deviations are caused by additional radiative corrections resulting from the additional scalars and pseudoscalars in the computation of self energy amplitudes of the SM gauge bosons. The electroweak precision measurements  estimate the deviations in the precision observables as  ~\cite{Workman:2022ynf}  
\be
\Delta S = - 0.02 \pm 0.10, \,\,
\Delta T = 0.03 \pm 0.12, \,\, 
\Delta U = 0.01 \pm 0.11  \label{eq:STU-values}
\ee

\par Defining $\Delta \mw =\mw - \mwsm$ and approximating $\Delta \mw^2  \simeq  2 \, \mwsm  \Delta \mw$, the discrepancy between the SM prediction and experimental value of $W$ mass can be computed using the relation
\be 
\Delta \mw = \frac{ \alpha\, \mwsm }{2(c_w^2 - s_w^2)} \left( -  \frac{1}{2}  \Delta S + c_w^2 \Delta T + \frac{c_w^2 - s_w^2}{4 s_w^2}  \Delta U \right) .
\label{eq:deltamw}
\ee
Since, the contribution from $\Delta U$  is small, henceforth  we consider only the corrections from $\Delta S$  and $\Delta T$ to   $\Delta \mw $.

\par We compute the  deviations $\Delta S$ and $\Delta T$  in  our model  at one loop level coming from  scalars and pseudo scalars $h_i,\, P^0,\,A^0 $. The  explicit expressions for the same are given in the appendix~\ref{app:STU}. The equation~\eqref{eq:deltamw} can then be solved  iteratively to determine the prediction of \mw in our model. 
\onecolumngrid
\begin{center}
	\begin{figure}[htb]
		\subfloat[{ \em{ $\theta_{23} = 30^\circ$, $R_P = \frac{m_{_{P^0}}}{m_{_{A^0}} }= 0.5$ }}\label{fig:contour-r-halfa}]{%
			\includegraphics[width=.49\linewidth]{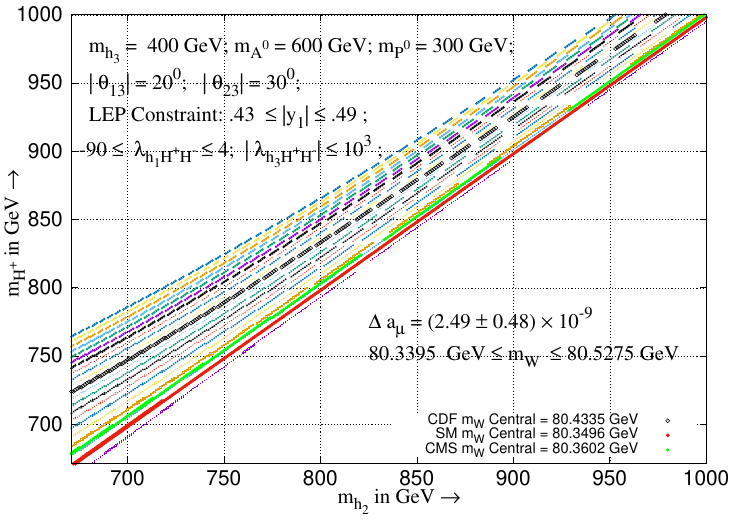}
		}\hfill%
		\subfloat[{ \em{ $\theta_{23} = 30^\circ$, $R_P = \frac{m_{_{P^0}}}{m_{_{A^0}} }= 1$ }}\label{fig:contour-r1}]{%
			\includegraphics[width=.49\linewidth]{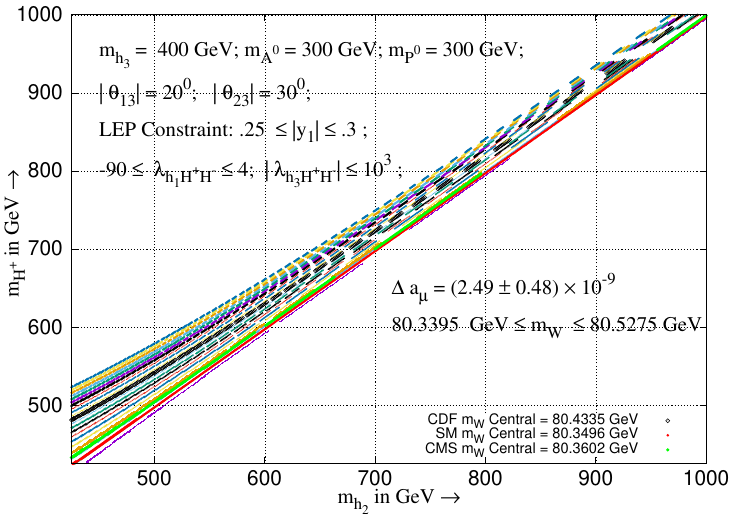}
		}\hfill%
		\subfloat[{ \em{ $\theta_{23} = 45^\circ$, $R_P = \frac{m_{_{P^0}}}{m_{_{A^0}} }= 0.5$}}\label{fig:contour-r-halfb}]{%
			\includegraphics[width=.49\linewidth]{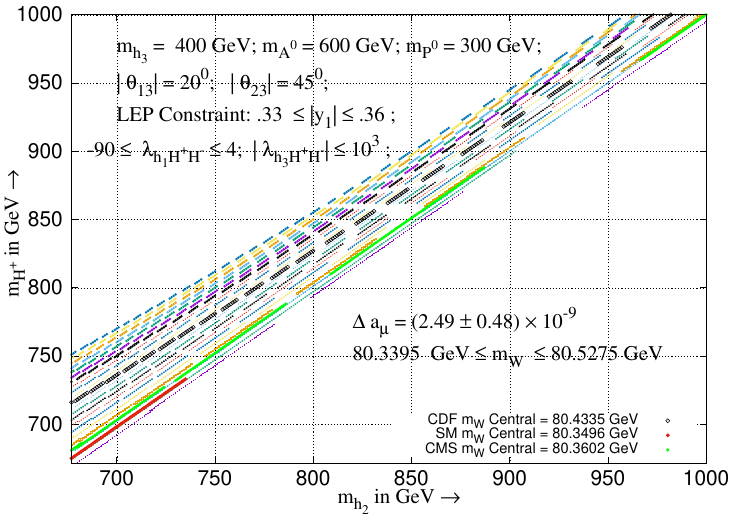}
		}\hfill%
		\subfloat[{ \em{ $\theta_{23} = 45^\circ$, $R_P = \frac{m_{_{P^0}}}{m_{_{A^0}} }= 2$ }}\label{fig:contour-r2}]{%
			\includegraphics[width=.49\linewidth]{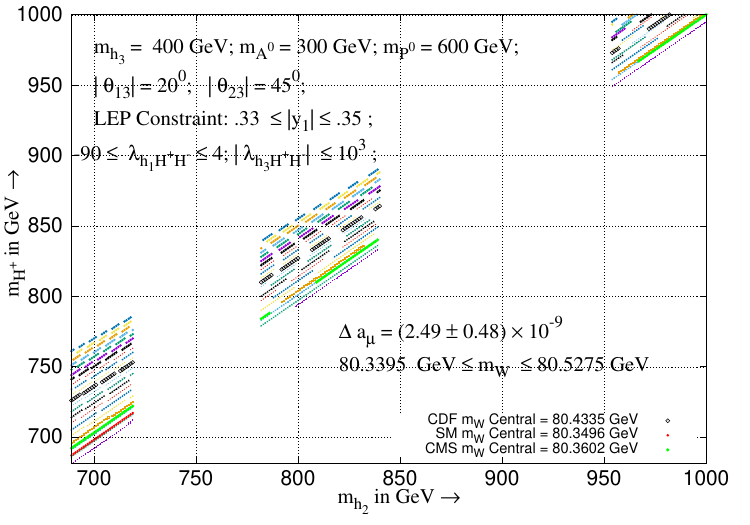}
		}\hfill%
		\caption[nooneline]{\justifying \em{This figure exhibits the allowed parameter space in the $m_{_{h_2}} -m_{H^\pm}$ plane for the parameter sets (a) $\theta_{23} = 30^\circ$, $R_P = \frac{m_{_{P^0}}}{m_{_{A^0}} }= 0.5$ , (b) $\theta_{23} = 30^\circ$, $R_P = \frac{m_{_{P^0}}}{m_{_{A^0}} }= 1$, (c) $\theta_{23} = 45^\circ$, $R_P = \frac{m_{_{P^0}}}{m_{_{A^0}} }= 0.5$  and (d) $\theta_{23} = 45^\circ$, $R_P = \frac{m_{_{P^0}}}{m_{_{A^0}} }= 2$. In each panel, the loci of points in a given color depict a contour satisfying simultaneously  (i) LEP and partial Higgs decay width constraints from LHC,  (ii) a specific value of $\mw = \mwcdf + n \sigma^{\rm CDF} $, with $n \in [-10, 10]$,  and (iii) muon anomalous  magnetic moment in the range  $[2.01 \, :  \,2.97]\times 10^{-9}  $  (1$\sigma$ band of $\Delta a_\mu $). The lowest (uppermost) contour   corresponds to $n=-10$ ($n=10$). The loci of red, green and black points correspond to the central values of $ \mwsm$, $\mwcms$ and $ \mwcdf$, respectively. }}
	\label{fig:contour}
	\end{figure}
\end{center}
\twocolumngrid
\section{The Anomalies}
\label{sec:results}
In this section we demonstrate  the simultaneous explanation of the twin anomalies while satisfying all the constraints discussed so far. Our numerical analysis algorithm is designed as follows:
\begin{itemize}
	\item Based on the LHC constraint on the partial decay width of Higgs to $W\, W^\star$ and identifying the lightest scalar in the spectrum to be $m_{h_1}$ = 125 GeV, we fix the CP-even mixing angle  $\left\vert\theta_{13}\right\vert=20^\circ$. 
	\item As discussed in section \ref{sec:constraints}, we divide the parameter space into three regions of pseudoscalar mass ratios: $R_P=m_{P^0}/m_{A^0} \equiv$  0.5, 1, and 2. Each such region is further investigated for three choices of CP-Odd mixing angle $\theta_{23}\equiv$ $30^\circ$, $45^\circ$, and $60^\circ$.
	\item In general all scalar and pseudoscalar masses are varied between 200 GeV and 1 TeV. However, in accordance with equations \eqref{eq:mh2a} and \eqref{eq:mh2a},  $m_{h_2}$ and $m_{H^+}$ are varied in range $\sqrt{m_{A^0}^2+m_{P^0}^2 }< m_{_{h_2}}, m_{{H^\pm}} \le $ 1 TeV. 
	For the purpose of demonstration and paucity of space, we choose specific mass combinations for  ($m_{A^0},\, m_{P^0} $):  ($600,\, 300$) GeV, ($300,\, 300$) GeV, and ($300,\, 600$) GeV corresponding to $R_P=m_{P^0}/m_{A^0} \equiv$  0.5, 1, and 2 respectively.  
	\item  The magnitude of the Yukawa coupling  is kept below the perturbative limit,  $\left\vert y_1\right\vert \le \sqrt{4\,\pi}$ and is strongly constrained from the LEP data.  
	The triple scalar coupling $\lambda_{h_1 H^+H^-}$ is varied in the allowed range for a given value of \mhpm while $|\lambda_{h_3 H^+H^-}|$ is probed in the range 0 to
		$ 10^3$. 
	\item Next, we scan the constrained parameter hyperspace  to  search for simultaneous solution for  $W$ mass lying in the range $[80.3395 : 80.5275]\,$ and the anomalous magnetic moment of muon  lying within one sigma band $\Delta a_\mu \in [2.01 :  2.97]\times 10^{-9}$~\cite{Muong-2:2023cdq} given in equation \eqref{eq:delamu}. The specified range of  $\mw$ is chosen in order to include $\mwcms$~\cite{CMS:2024nau},  $\mwsm$~\cite{deBlas:2022hdk}  as well as $\mwcdf$~\cite{CDF:2022hxs}. 
\end{itemize}

\par Following the analysis, figure~\ref{fig:contour} illustrates this 
allowed parameter space in the $m_{_{h_2}} -m_{H^\pm}$ 
plane for various combinations of $\theta_{23}$ and $R_P$ that 
satisfy the one sigma permissible range for $\Delta a_\mu$. The 
lower limits of $m_{_{h_2}} $ and $m_{_{H^\pm}} $ in these plots 
are set by the constraint~\eqref{eq:lowlim-mh2mhp}. The contour 
satisfying a specific value of $\mw = \mwcdf + n \sigma^{\rm CDF} $, with $n \in [10, 10]$, is represented by the loci of points 
	in a given color. The lowest (uppermost) contour   
	corresponds to $n=-10$ ($n=10$). The loci of red, green and black points 
	correspond to the central values of $ \mwsm$ , $\mwcms$ and $\mwcdf$ 
	respectively\footnote{While this work was under 
		review, the new result of $W$-mass measurement by CMS 
		collaboration was announced~\cite{CMS:2024nau}.  Although we have centered the contours around  the CDF value pof \mw, it may 
		 be noted that there is enough parameter space that favors the the CMS central value   as well as SM global fit values of \mw  as shown in figure~\ref{fig:contour}. }.  The 
		choice of $\left\vert y_1\right\vert$ for a given set of scalar 
		and pseudoscalar masses are essentially dictated by the LEP 
		constraint and hence varies within a narrow range as 
		mentioned  in the legend. We  make some important 
		observations on the contour plots based on the model 
		analysis:  
\begin{itemize}
	\item Since the Yukawa coupling of leptons with $A^0$ and $P^0$ is proportional to $\cos{\theta_{23}}$ and $\sin{ \theta_{23}}$ respectively, the behaviour of the contour plots for $\theta_{23} = 30^\circ$ and $R_P = 0.5$ is very similar to the case with $\theta_{23} = 60^\circ$ and $R_P = 2$. Hence we show plot for only one of them in the figure~\ref{fig:contour-r-halfa}.
	\par On the similar note, for cases where the mass ratio $R_P$ is unity,   the LEP constraints and the  value of $\Delta a_\mu$ become independent of the mixing angle $\theta_{23}$, and hence, similar patterns are obtained in the contour plots for all $\theta_{23}$. We have therefore  depicted only one of them for $\theta_{23}=30^\circ$ in the  figure~\ref{fig:contour-r1}. 
	
	\item No viable solution for \mw in the required range is found for $R_P=2(0.5)$, at fixed $m_{h_3}$ = 400 GeV keeping all other parameters constant, for $\theta_{23} = 30^\circ $($60^\circ$). However, given a lower (or higher) value of $m_{h_3}$, the solution does exists. This is also evident from the $m_{h_3}$ color density  plot in  figures~\ref{fig:damu-mwb}(\ref{fig:damu-mwe}).
	\item The long discontinuities of loci of points  in the contour plots of figure~\ref{fig:contour} indicate the noncompliance  of the model parameters  to accommodate measured values of  both  observables simultaneously in the required range.
\end{itemize}

\par Finally, we exhibit the sensitivity of model parameter space through the $m_{h_3}$ color density maps in the  $ \Delta a_{\mu} - \mw$ plane in figure~\ref{fig:damu-mw}  for  different combinations of $R_P$ and  $\theta_{23} $. The black horizontal  lines corresponds to  1$\sigma$ band  of  \mwcdf given by~\eqref{eq:wmass-cdf} while the red horizontal line corresponds to the  central \mwsm  value~\eqref{eq:wmass-SM}.
We also depict the recently announced value of \mw by CMS~\eqref{eq:wmass-cms}  by green horizontal line. Similarly, the vertical blue line corresponds to the central value of $\Delta a_\mu$ given by~\eqref{eq:delamu}. A couple of observations from the figure~\ref{fig:damu-mw} are given below:
\begin{itemize} 
	\item  For a given $R_P$, 
	lower values of   $m_{_{h_3}} $ are favored for lower values of  mixing angle $\theta_{23}$. Similarly, for a given value of $\theta_{23}$,  lower values of   $m_{_{h_3}} $ are favored for higher  $R_P$. 
	\item  For $\theta_{23} =30^\circ$ and $R_P =2, m_{A^0}=300\,$GeV, the common parameter space allowed by \mw value favors $\Delta a_\mu$  in the lower half of $1\sigma$  band while for $\theta_{23} =60^\circ$ and $R_P =0.5, m_{A^0}= 600\,$GeV, the  parameter space allowed by \mw value favors $\Delta a_\mu$  in the upper half of $1\sigma$  band. This can be inferred from figures~\ref{fig:damu-mwb} and \ref{fig:damu-mwe}.
\end{itemize}


\par
Thus, we see that a fairly large mutually exclusive regions in the  parameter space of the model  are available that accommodate the  CDF, SM and CMS  values of $W$-boson mass while simultaneously
solving the anomaly of $\Delta a_\mu$.  

\section{Summary}
\label{sec:summary}
In this article, we have considered a minimal extension of the inert 2HDM with the inclusion of a $Z_2$ odd $SU(2)$ complex scalar singlet  to explain the deviations of the recent measurements of the muon anomalous magnetic moment and  the $W$-boson mass from their SM predictions. Implementing the stability and minimization conditions on the scalar potential, we have parameterized the model in terms of three neutral CP-even and two CP-odd scalar masses,  one charged Higgs mass, one mixing angle each for the CP-even and CP-odd pair of scalars,  Yukawa  coupling and scalar triple couplings of charged scalar. 

\par We identify the lightest scalar $h_1$ of the spectrum of this extended model with SM-like Higgs ($m_{h_1}$ = 125 GeV) observed at LHC. The model is then constrained by the recent measurements of the partial decay widths of Higgs to gauge bosons at the LHC that fix the CP-even mixing angle $\theta_{13}\approx 20^\circ$. The average experimental values of signal strengths $\mu_{{}_{\gamma \gamma}}=1.10\pm0.07$ and $\mu_{{}_{W W^\star}}=1.00 \pm 0.08$~\cite {Workman:2022ynf} provide the allowed range for neutral scalar triple coupling  $\lambda_{{h_1 H^+ H^-}} $ with the charged Higgs. Further, the existing LEP data for $\sigma (e^+\ e^- \to \mu^+\ \mu^-)=3.072 \pm 0.108 \pm 0.018\, { \rm pb }$~\cite{ALEPH:2013dgf} is used to constrain the relation between the Yukawa couplings and the masses of the scalar and pseudoscalars as stated in  equation~\eqref{xsec:mumu}.

\par We then compute the contribution of the model to the anomalous magnetic moment of the charged lepton, $\Delta a_\mu$, at the one loop level arising from the Feynman diagrams due to the exchange of the neutral scalar, pseudoscalar, and charged scalars as given in \eqref{eq:MDMoneloop}. The contribution of the dominant Bar-Zee diagrams at the two-loop level  presented  in equation \eqref{eq:MDM-2loop} is also included in the computation of $\Delta a_\mu$. 

\par Next, we calculate the contribution to the precision observables $\Delta S$ and $\Delta T$  at the one loop level from the scalars and pseudoscalars in the extended I2HDM as given in the appendix~\ref{app:STU}.  This deviation of the precision variables from the SM  prediction is fed into the nonlinear relation for the $W-$boson mass in equation \eqref{eq:deltamw}.
We then solve this nonlinear equation iteratively by varying the model's parameters to compute the contribution to $W$-boson mass in the model.

\par The constrained model is systematically scanned and analysed to accommodate both experimental observations simultaneously. For simplicity and brevity, the analysis is reported  for three pseudoscalar  mass combinations ($ m_{A^0}, m_{P^0}$) $\equiv$ (300,300)~GeV, (300, 600)~GeV, and (600,300)~GeV and three choices of the pseudoscalar mixing angle $\theta_{23}\equiv 30^\circ, 45^\circ$, and $60^\circ$.  The $m_{h_2}$, $m_{h_3}$, and $m_{H^+} $ are varied up to 1 TeV, while the lower limits for $m_{h_2}$ and $m_{H^+}$ are fixed by the equations \eqref{eq:mh2a} and \eqref{eq:mhpm}, respectively. Maintaining the unitarity of Yukawa couplings, the coupling $|y_1|$ is varied in range $|y_1|< \sqrt{4\pi}$. The allowed values of $|y_1|$  are fixed from the LEP data and the one sigma range for $ \Delta a_{\mu} = \left(249 \pm 48  \right) \times 10^{-11}$ \cite{Muong-2:2023cdq}.

\par
\begin{center}
	\begin{figure}[h!]
		\subfloat[{ \em{ $\theta_{23}= 30^\circ$,  $R_P = 0.5$, $m_{_{A^0}} = 600\,$GeV   }}\label{fig:damu-mwa}]{\includegraphics[width=.49\linewidth]{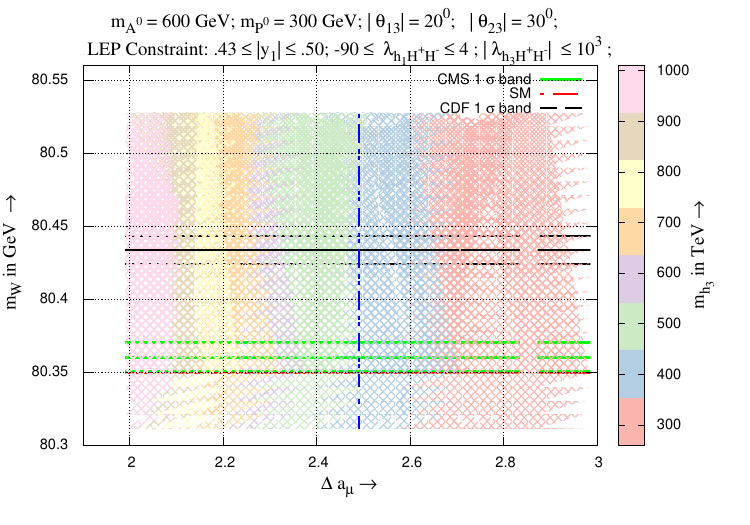}}
		\subfloat[{ \em{ $\theta_{23}= 60^\circ$,  $R_P = 0.5$, $m_{_{A^0}} = 600\,$GeV   }}\label{fig:damu-mwb}]
		{\includegraphics[width=.49\linewidth]{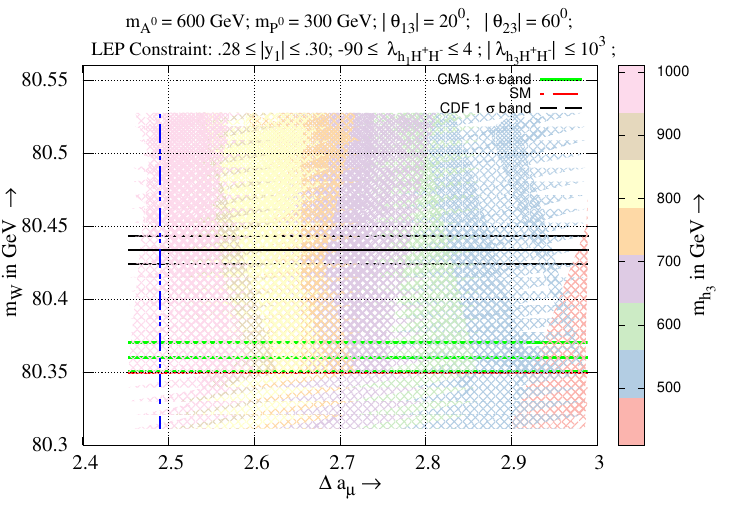}}\\
		\subfloat[{\em{$\theta_{23}= 30^\circ$,  $R_P = 1$, $m_{_{A^0}} = 300\,$GeV }}\label{fig:damu-mwc}]{\includegraphics[width=.49\linewidth]{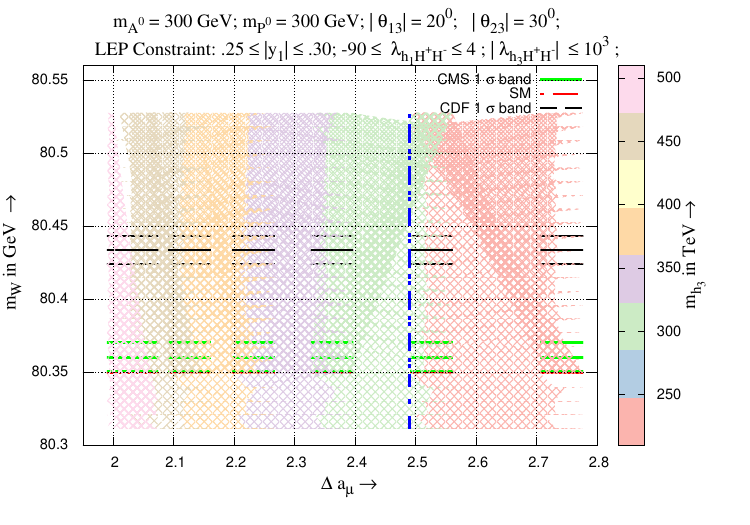}}
		\subfloat[{\em{$\theta_{23}= 60^\circ$,  $R_P = 1$, $m_{_{A^0}} = 300\,$GeV   }}\label{fig:damu-mwd}]{\includegraphics[width=.49\linewidth]{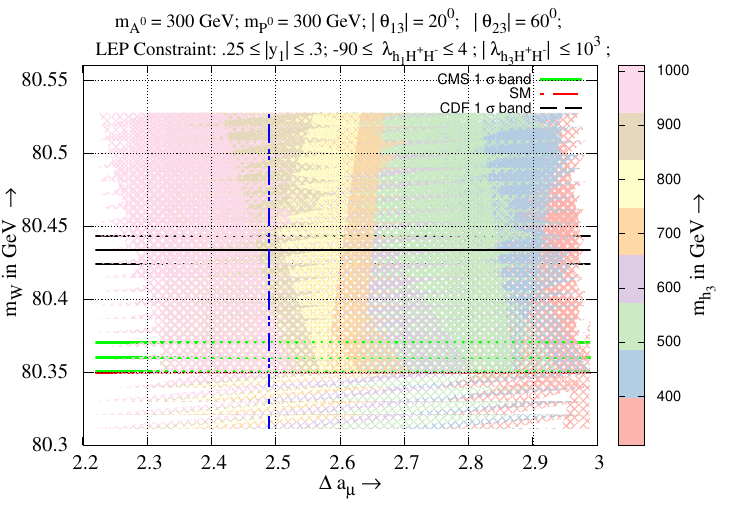}}\\
		\subfloat[{\em{$\theta_{23}= 30^\circ$,  $R_P = 2$, $m_{_{A^0}} = 300\,$GeV   }}\label{fig:damu-mwe}]{\includegraphics[width=.49\linewidth]{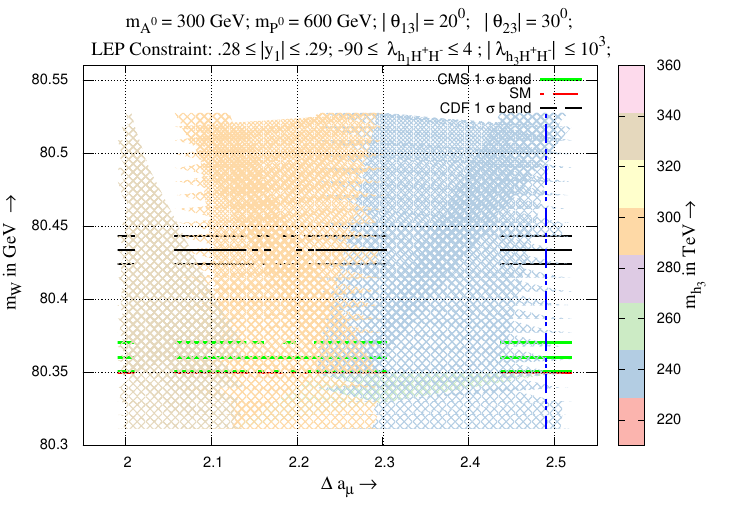}}
		\subfloat[{\em{$\theta_{23}= 60^\circ$,  $R_P = 2$, $m_{_{A^0}} = 300\,$GeV   }}\label{fig:damu-mwf}]{\includegraphics[width=.49\linewidth]{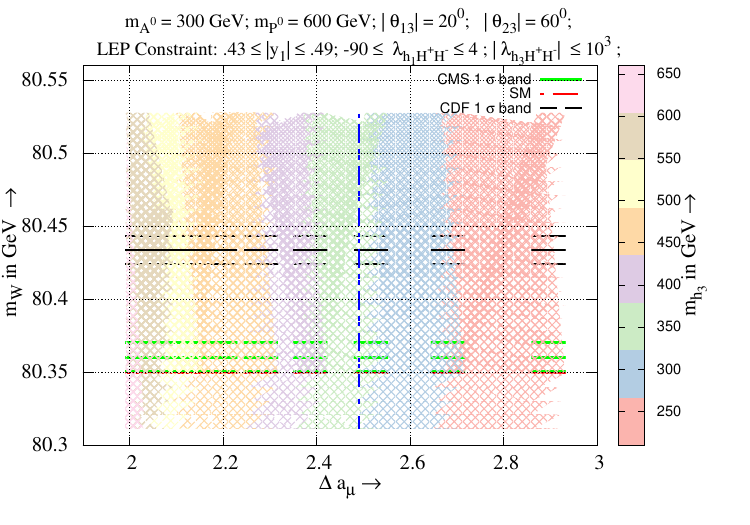}}
		\caption[nooneline]{\justifying \em{Model prediction for twin anomalies through the $m_{h_3}$ color density maps in the $ \Delta a_{\mu} - \mw$  plane corresponding to  (a) $\theta_{23}= 30^\circ$,  $R_P = 0.5$, $m_{_{A^0}} = 600\,$GeV , (b) $\theta_{23}= 60^\circ$,  $R_P = 0.5$, $m_{_{A^0}} = 600\,$GeV  , (c) $\theta_{23}= 30^\circ$,  $R_P = 1$, $m_{_{A^0}} = 300\,$GeV, (d) $\theta_{23}= 60^\circ$,  $R_P = 1$, $m_{_{A^0}} = 300\,$GeV  (e) $\theta_{23}= 30^\circ$,  $R_P = 2$, $m_{_{A^0}} = 300\,$GeV and  (f) $\theta_{23}= 60^\circ$,  $R_P = 2$, $m_{_{A^0}} = 300\,$GeV. The density plot satisfy LEP limits and partial Higgs decay width constraint from LHC. The black horizontal dashed lines correspond to \mwcdf 1$\sigma$ band given by~\eqref{eq:wmass-cdf} while the red horizontal dashed line corresponds to \mwsm predicted value~\eqref{eq:wmass-SM}. The recently announced value of \mw by CMS~\eqref{eq:wmass-cms}  is shown by green  horizontal line. The vertical blue line corresponds to the central value of $\Delta a_\mu$ given by~\eqref{eq:delamu}. The values of  $\Delta a_\mu$ are shown only in the 1$\sigma$ band, i.e. in the range $[2.01:2.97] \times 10^9$.”} }
		\label{fig:damu-mw}
	\end{figure}
\end{center}

 Our analysis can be summarised from the four panels in figure~\ref{fig:contour}, where each panel consists of 22 contour plots in the $m_{_{h_2}} -m_{H^\pm}$ plane  for various combinations of $\theta_{23}$ and ratio $R_P = m_{P^0}/m_{A^0}$. The contours  correspond to various \mw values in the range $[80.3395 : 80.5275]$ including the three central values, namely, \mwcdf, \mwcms and \mwsm 
 and simultaneously satisfy the one sigma permissible range for $ \Delta a_{\mu}$ at fixed $m_{h_3} = 400 \,$GeV. The observations are further reinforced  by depicting the allowed common parameter space in the color density plots for $m_{h_3}$ in the $ \Delta a_{\mu} - \mw$ plane in figure~\ref{fig:damu-mw}  for  different combinations of $R_P$ and  $\theta_{23} $.

\par  Thus, the LEP and LHC data-constrained parameter hyperspaces of the said model accommodate recent observations of both $\Delta a_\mu$ and  \mw. Although we have worked with a restricted parameter space, the simultaneous solution space of the parameters is, however, fairly large and also spans over other choices of the mass combinations for pseudoscalars with the mixing angle $20^\circ\le\left\vert\theta_{23}\right\vert \le 80^\circ$. 
\acknowledgments
We acknowledge the partial financial support from SERB grant CRG/2018/004889. MD would like to thank Inter University Center for Astronomy and Astrophysics (IUCAA), Pune for hospitality while part of this work was completed.

\onecolumngrid
\twocolumngrid
%

\appendix
\label{sec:appendix}
\begin{widetext}
\section{Scalar Couplings in terms of Mass Parameters}
\label{app:coup-mass}
The minimisation of the potential leads to following relations:
\bea
m^2_{11} &=& \l_1 \vsm^2 + \l_{11} v^2_s, \\
m^2_{33} &=& \l_8 v^2_s + \l_{11} \vsm^2.
\eea
Further, the mass relations given by equations, can be combined to give the following equations relating the couplings appearing in the scalar potential~\eqref{eq:scalarpot} with the physical mass parameters:
\begin{eqnarray}
	\lambda_3 &=&  \frac{1}{v_{\rm SM}^2} \left[2\,m^2_{H^\pm} + m^2_{22} -\lambda_{13}\,\, v_s^2\right]  \label{eq:lam3} \\
	\lambda_4 &=& \frac{1}{v_{\rm SM}^2} \left[m_{h_2}^2+ m_{A^0}^2+ m_{P^0}^2-2\,m_{H^\pm}^2\right]. \label{eq:lam4} \\
	\lambda_5  &=& \frac{1}{v_{\rm SM}^2}\left[m_{h_2}^2-m_{A^0}^2- m_{P^0}^2\right] \label{eq:lam5}\\
	\lambda_8 &=& \frac{1}{v_{s}^2}\left[m_{h_1}^2 + m_{h_3}^2 -\lambda_1\, v_{\rm SM}^2\right] \label{eq:lam8}\\
	\lambda_{11} &=& \frac{1}{v_{\rm SM}\,v_{s}}\left(\lambda_1\, v_{\rm SM}^2 -\lambda_8\, v_{s}^2\right)\tan\left(2\,\theta_{13}\right) \label{eq:lam11}\\
	\kappa &=&  -\,\frac{1}{2\,\sqrt{2} \, \vsm  }\left(m^2_{P^0}+m^2_{A^0}\right)\tan\left(2\,\theta_{23}\right)\label{eq:kap}
\end{eqnarray}

Thus, considering  VEVs $v_{\rm SM}$ and $v_{s}$, mixing angles $\theta_{13}$ and $\theta_{23}$, coupling $\lambda_{13}$ and  masses 
$m_{22}^2,\,\, m_{h_1}^2, \,\, m_{h_2}^2, \,\, m_{h_3}^2, \,\, m_{H^\pm}^2, \,\, m_{A^0}^2, \,\, {\rm and} \,\, m_{P^0}^2$ to be the free parameters, we can express
$m_{11}^2,\,\, m_{33}^2,\,\, \lambda_3,\,\, \lambda_4,\,\, \lambda_5 \,\, \lambda_8, \,\,\lambda_{11} \,\, {\rm and} \,\, \kappa$ in terms of the above free parameters.
	\section{Definition of Loop Form Factors}
	\label{app:formfactors}
	The loop amplitudes used in equations \eqref{hsmtoaa} and  \eqref{zetaaa} are expressed in terms  of dimensionless parameter $\tau$, which is essentially function of the ratios of mass squared  of physical scalars, pseudoscalars, gauge bosons and fermions. 
	\begin{subequations}
		\begin{eqnarray}
			{\cal M}^{\gamma\gamma}_0(\tau)&=&-\tau[1-\tau f(\tau)]\\
			{\cal M}^{\gamma\gamma}_{1/2}(\tau)&=&2\tau[1+(1-\tau)f(\tau)],\\
			{\cal M}^{\gamma\gamma}_1(\tau)&=&-[2+3\tau+3\tau(2-\tau)f(\tau)]
			\nn\\
			\label{eq:hzgaform}
		\end{eqnarray}
	\end{subequations}
	where
	\vskip -1cm
	\bea
	f(\tau)=
	\left\{ 
	\begin{array}{cc}
		\arcsin^2\big(\frac{1}{ \sqrt{\tau} }\big) & \textrm{for } \, \tau\geqslant 1,\\
		-\frac{1}{4}\Big[\log\Big(\frac{1+\sqrt{1-\tau}}{1-\sqrt{1-\tau}}\Big)-i\pi\Big]^2  & \textrm{for } \,\tau<1.\\
	\end{array}
	\right.
	\eea
	\section{One loop and two loop functions for MDM}
	\label{app:MDMLoopFunc}
	The integrals required to compute the one loop contribution to the muon magnetic moment  of leptons \eqref{eq:MDMoneloop} are given by
	\begin{subequations}
		\bea
		{\cal I}_1(r^2)&=&\int_0^1 dx\ \frac{(1+x)(1-x)^2}{(1-x)^2\ r^2+ x}\label{i1eq}\\
		{\cal I}_2(r^2)&=&\int_0^1 dx\ \frac{-(1-x)^3}{(1-x)^2\ r^2+ x},\label{i2eq}\\
		{\cal I}_3(r^2)&=& \int_0^1 dx\ \frac{-x(1-x)}{1- (1-x)\  r^2}\label{i3eq}
		\eea
	\end{subequations}
	with $r=\frac{m_l}{m_{s_i}},$ and $s_i= h_1,\ h_2,\ h_3,\ A^0,\  P^0.$
	
	\par The integrals contributing to the muon magnetic moment of leptons at two loop level given in equation \eqref{eq:MDM-2loop}  are defined as
	\begin{subequations}
		\begin{eqnarray}
			f(r^2) &=& \frac{r^2}{2}\, \int_0^{1} dx\ \frac{1\ -\ 2x(1-x)}{x(1-x)\ -\ r^2}\
			\ln \left[\,\frac{x(1-x)}{r^2}\,\right]\label{2loopint1}\\
			\tilde{f}(r^2) &=& \int_0^{1} dx\ \frac{x(1-x)}{r^2 -\ x(1-x)}\
			\ln \left[\,\frac{x(1-x)}{r^2}\,\right] \label{2loopint3}
		\end{eqnarray}
	\end{subequations}
	\section{The Oblique Parameters}
	\label{app:STU}
		The precision observables derived from the radiative corrections of the gauge Boson propagator are essentially the two point vacuum polarization tensor functions of 
		$\Pi_{ij}^{\mu\nu} (q^2)$, $q^2$ is the four-momentum of the vector boson 
		($V = W, Z or \gamma$). Following the
		prescription of the reference~\cite{Haber:1993wf} the vacuum polarization tensor functions corresponding to pair of gauge Bosons $V_i , V_j$ can be written as
		\begin{subequations}
			\bea
			i\Pi^{\mu\nu}_{ij}(q)&=&ig^{\mu\nu}A_{ij}(q^2)+iq^\mu q^\nu
			B_{ij}(q^2) \quad ; \quad\quad
			A_{ij}(q^2)=A_{ij}(0)+q^2F_{ij}(q^2) \label{ewp}
			\eea
		\end{subequations}
	
	The oblique parameters  are defined as:
	\begin{subequations}
		\bea
		S&\equiv& \frac{1}{g^2} \left(16\pi \cos\theta_{W}^2\right)\left[F_{ZZ}(m_Z^2)
		-F_{\gamma\gamma}(m_Z^2) +\left(\frac{2\sin\theta_{W}^2-1}{ \sin\theta_{W}
			\cos\theta_{W}}\right)F_{Z\gamma}(m_Z^2)\right]\label{sdef}\\
		T&\equiv & \frac{1}{\alpha_{em}}\left[\frac{A_{WW}(0)}{ m_W^2}-{A_{ZZ}(0)\over
			m_Z^2}\right] \label{tdef}\\
		U&\equiv& \frac{1}{g^2}\left( 16\pi\right)\left[ F_{WW}(m_W^2)-F_{\gamma\gamma}(m_W^2) -
		{\cos\theta_{W}\over \sin\theta_{W}}F_{Z\gamma}(m_W^2)\right]  - \, S \,.\label{udef}
		\eea
	\end{subequations}
	$\alpha_{em}$ being the fine structure constant. It is worthwhile to mention that although $A_{ij}(0)$ and $F_{ij}$ are divergent by themselves but the total divergence associated with each precision parameter in equations \eqref{sdef},  \eqref{tdef} and \eqref{udef} vanish on taking into account a gauge invariant set of  one loop diagrams contributing  for a given pair of gauge Bosons.  
	The additional contribution to the oblique parameters (apart from SM) in our model can be computed to give
	\begin{subequations}
		\bea
		\hspace*{-1cm} \Delta S&=& \frac{G_F\,\alpha_{em}^{-1}}{2\sqrt{2}\, \pi^2} \,\sin^2\left(2\,\theta_W\right)\,\Bigg[\sin^2\theta_{13} \bigg\{m_Z^2\bigg( \mathcal{B}_0(m_Z^2;m_Z^2,m_{h_1}^2)-  \mathcal{B}_0(m_Z^2;m_Z^2,m_{h_3}^2)   \bigg) 
		+  \mathcal{B}_{22}(m_Z^2;m_Z^2,m_{h_3}^2)
		\nn\\&&
		- \mathcal{B}_{22}(m_Z^2;m_Z^2,m_{h_1}^2)\bigg\}
		+ \cos^2\theta_{23} \mathcal{B}_{22}(m_Z^2;m_{h_2}^2,m_{A^0}^2) + \sin^2\theta_{23} \mathcal{B}_{22}(m_Z^2;m_{h_2}^2,m_{P^0}^2) - \mathcal{B}_{22}(m_Z^2;m_{H^\pm}^{2},m_{H^\pm}^{2})  \Bigg]\nn\\
		\label{eq:deltaS}
		\eea 
		where
		\vskip -0.5cm
		\bea
		\mathcal{B}_{22}(q^2;m_1^2,m_2^2) & = & B_{22}(q^2;m_1^2,m_2^2)- B_{22}(0;m_1^2,m_2^2)\\
		\mathcal{B}_{0}(q^2;m_1^2,m_2^2) & = & B_{0}(q^2;m_1^2,m_2^2)- B_{0}(0;m_1^2,m_2^2)
		\eea
	\end{subequations}
	\bea
	\Delta T&=& \frac{G_F\,\alpha_{em}^{-1}}{2\sqrt{2}\, \pi^2}\Bigg[ \sin^2\theta_{13} \Bigg\{m_W^2 \bigg(B_0(0;m_W^2,m_{h_1}^2)- B_0(0;m_W^2,m_{h_3}^2) \bigg)
	-m_Z^2 \bigg( B_0(0;m_Z^2,m_{h_1}^2) - B_0(0;m_Z^2,m_{h_3}^2) \bigg) 
	\nn\\&&
	+  B_{22}(0;m_W^2,m_{h_3}^2) - B_{22}(0;m_W^2,m_{h_1}^2)
	+   B_{22}(0;m_Z^2,m_{h_1}^2)-  B_{22}(0;m_Z^2,m_{h_3}^2) \Bigg\}
	\nn\\ &&
	- \frac{1}{2}\, A_0(m^2_{H^\pm})+ B_{22}(0;\mhpm^2,m_{h_2}^2)
	+\cos^2\theta_{23} \bigg( B_{22}(0;\mhpm^2,m_{A^0}^2)-B_{22}(0;m_{h_2}^2,m_{A^0}^2)\bigg)   \nn\\
	&&+ \sin^2\theta_{23} \bigg(B_{22}(0;\mhpm^2,m_{P^0}^2) - B_{22}(0;m_{h_2}^2,m_{P^0}^2)\bigg) 
	\Bigg]
	\eea

	The Veltman Passarino Loop Integrals $A_0,\, B_0,\, B_{22}$  in the above expressions are defined as 
	\begin{subequations}
		\bea
		A_0(m^2)&=& m^2\ \left( \Delta + 1 - \ln m^2  \right),\\
		B_0(q^2;m_1^2,m_2^2)&=&\Delta-\int_0^1\,dx\,\ln(X-i\epsilon)\\
		B_{22}(q^2;m_1^2,m_2^2)&=&\frac{1}{4}(\Delta+1)\left[m_1^2+m_2^2
		-\frac{1}{3} q^2\right]-\frac{1}{2}\int_0^1\,dx\,X\ln(X-i\epsilon)
		\eea
	\end{subequations}
	where $X\equiv m_1^2x + m_2^2(1-x) - q^2 x(1-x)$ and $\Delta\equiv{2\over 4-d}+\ln(4\pi)+\gamma_E$ in $d$ space-time dimensions. For the Feynman rules and Feynman diagrams involved in the computation of vacuum polarisation functions for
	$\Delta S$  and $\Delta T$, one is referred to the reference~\cite{Bharadwaj:2021tgp}.
	
\end{widetext}
\bibliography{refbddg_final}
\end{document}